\documentclass[11pt,a4paper]{article}
\pdfoutput=1
\usepackage{jheppub}
\allowdisplaybreaks
\usepackage{color}
\usepackage{amsmath}
\usepackage{graphicx}
\usepackage{esint}
\usepackage{enumerate}
\usepackage{slashed}
\usepackage[english]{babel}
\usepackage{hyperref}
\usepackage{afterpage}
\newcommand{\blue}{\textcolor{blue}}

\usepackage{babel}
\begin{document}

\title{MUonE sensitivity to new physics explanations of the muon anomalous magnetic moment}

\author[a]{P. S. Bhupal Dev,}
\author[b]{Werner Rodejohann,}
\author[b]{Xun-Jie Xu,}
\author[a]{Yongchao Zhang}
\affiliation[a]{Department of Physics and McDonnell Center for the Space Sciences,  Washington University, \\
St.\ Louis, MO 63130, USA}
\affiliation[b]{Max-Planck-Institut f\"ur Kernphysik, Saupfercheckweg 1,  69117 Heidelberg,  Germany}

\abstract{
The MUonE experiment aims at a precision measurement of the hadronic vacuum polarization contribution to the muon $g-2$, via elastic muon-electron scattering. Since the current muon $g-2$ anomaly hints at the potential existence of new physics (NP) related to the muon, the question then arises as to whether the measurement of hadronic vacuum polarization in MUonE could be affected by the same NP as well. In this work, we address this question by investigating a variety of NP explanations of the muon $g-2$ anomaly via either vector or scalar mediators with either flavor-universal, non-universal or even flavor-violating couplings to electrons and muons. We derive the corresponding MUonE sensitivity in each case and find that the measurement of hadronic vacuum polarization at the MUonE is not vulnerable to any of these NP scenarios.
}


\maketitle

\section{Introduction}
The muon anomalous magnetic moment ($g-2$) is known to have a noteworthy discrepancy between the experimental observations and the Standard Model (SM) predictions for more than a decade. Currently, the experimental value of $a_{\mu}=(g-2)_\mu/2$ is~\cite{Bennett:2006fi, Tanabashi:2018oca}
\begin{equation}
a_{\mu}^{{\rm exp}} \ = \ (11659209.1\pm 6.3)\times10^{-10},\label{eq:a-exp-val}
\end{equation}
whereas the SM prediction is~\cite{Tanabashi:2018oca,Blum:2013xva}
\begin{equation}
a_{\mu}^{{\rm SM}} \ = \ (11659183.0\pm 4.8)\times10^{-10} \, , \label{eq:a-SM-val}
\end{equation}
so the difference between experiment and theory is currently at $3.3 \sigma$ level:
\begin{equation}
\Delta a_{\mu} \ \equiv \ a_{\mu}^{{\rm exp}}-a_{\mu}^{{\rm SM}} \ = \ (26.1\pm7.9)\times10^{-10} \,.
\label{eq:a_mu_val}
\end{equation}
This longstanding anomaly has motivated a plethora of theoretical explorations of possible new physics (NP) scenarios; for reviews, see e.g. Refs.~\cite{Jegerlehner:2009ry,Lindner:2016bgg}. Therefore, further efforts to either confirm  or eliminate the discrepancy with higher statistical confidence are of great significance. In the very near future, the Muon $g-2$ experiment at Fermilab is expected to publish their first result which will improve the current experimental precision by a factor of four~\cite{Grange:2015fou}, making the theoretical uncertainty  the dominant one in this theoretical-experimental discrepancy. A similar precision is also aimed at the future J-PARC experiment for measuring muon $g-2$~\cite{Abe:2019thb}.

On the theory front, the SM prediction for $a_\mu$ is generally divided into three parts, corresponding to the QED, electroweak (EW) and hadronic (had) contributions~\cite{Tanabashi:2018oca,Blum:2013xva}:
\begin{equation}
 a_\mu^\text{SM} \ = \ a_\mu^\text{QED} + a_\mu^\text{EW} + a_\mu^\text{had} \, .
\end{equation}
The QED and EW contributions are known to a very high accuracy~\cite{Gnendiger:2013pva, Aoyama:2017uqe}. So the theoretical uncertainty is currently dominated by the hadronic loop contributions~\cite{Davier:2017zfy, Keshavarzi:2018mgv,  Davier:2019can}, which are difficult to calculate precisely from first principles in the low-energy non-perturbative regime of QCD. Currently, the theoretical evaluation of the lowest-order hadronic contribution to $a_\mu^\text{SM}$~\cite{Davier:2019can},
\begin{equation}
a_{\mu}^{{\rm had}} \ = \ (693.9\pm 4.0)\times10^{-10},\label{eq:a-had}
\end{equation}
has an accuracy of 0.6\%, which is obtained from the R-ratio method. The state-of-the-art lattice QCD  techniques~\cite{Borsanyi:2017zdw, Blum:2018mom, Davies:2019efs, Gerardin:2019rua, Aubin:2019usy, Blum:2019ugy, Borsanyi:2020mff, Lehner:2020crt} determine $a_{\mu}^{{\rm had}}$ with larger uncertainties. Ref.~\cite{Borsanyi:2020mff} claims to have the smallest errors of all lattice calculations to date with $a_{\mu}^{\rm had}=(712.4\pm 4.5)\times 10^{-10}$, which is $3.1\sigma$ higher than the value in Eq.~\eqref{eq:a-had}. If this becames the true value, it would have resolved the muon $g-2$ anomaly. However, in this case the discrepancy between the R-ratio and lattice methods would still need to be explained~\cite{Lehner:2020crt}. Moreover, in Ref.~\cite{Crivellin:2020zul} it has been argued that adjusting $a_{\mu}^{\rm had}$ to the value suggested by Ref.~\cite{Borsanyi:2020mff} would actually shift the tension from  muon $g-2$  to the electroweak global fits.

To reduce the theoretical uncertainty, the MUonE experiment \cite{Abbiendi:2016xup} has been proposed, aiming at a new approach to determine the hadronic vacuum polarization more accurately.
Using a 150 GeV muon beam from CERN to scatter off atomic electrons of a low-$Z$ target (such as Beryllium layers), the MUonE experiment can achieve a very high-precision measurement of  elastic $\mu$--$e$ scattering at a QED-dominated momentum exchange of $q^2 =  {\cal O}(100 \ {\rm MeV})^2$.  By measuring the differential cross section ${\rm d}\sigma / {\rm d}T$ (with $T$ being the electron recoil energy), this low-energy high-precision measurement is very sensitive to the hadronic vacuum polarization of the photon mediator. With an average intensity of $1.3\times 10^7$ muon/s and two years of data taking, the MUonE experiment will be able to collect a total number of $3.7 \times 10^{12}$ $\mu$--$e$ scattering events and measure $a^{\rm had}_{\mu}$ with a statistical uncertainty of $\sim0.3$\%.

Considering, however, that there may be NP underlying the current muon $g-2$ anomaly, one may be concerned about whether and to what extent the potential existence of NP could affect the MUonE measurement of $a_{\mu}^{\rm had}$. On one hand, if the MUonE measurement of $a^{\rm had}_{\mu}$ shows a discrepancy with the theoretical prediction in the SM, one  would like to know whether this could be caused by the muon $g-2$ NP or is simply due to inaccurate QCD calculations.  On the other hand, as the MUonE center-of-mass energy is below GeV-scale, it is most sensitive to light NP degrees of freedom, such as a light scalar ($S$) or a light gauge boson ($Z'$). Given the high-precision measurements of electron and muon couplings in the SM~\cite{Tanabashi:2018oca}, if a heavy particle exists, its effects on MUonE is usually suppressed by its mass, unless it has a large coupling to the SM. For instance, a TeV-scale particle will need to have a coupling of order $4\pi$ to have the same effect on MUonE as a GeV-scale particle with coupling $10^{-2}$. Furthermore, the $\mu$--$e$ scattering receives tree-level contributions only if the new particle $X$ has {\it direct} couplings to electron and muon in the form of $X$--$e$--$e$, $X$--$\mu$--$\mu$ or $X$--$e$--$\mu$; otherwise, the NP contribution to $\mu$--$e$ scattering will be induced at loop level, even if the particle $X$ is a good candidate for the muon $g-2$ discrepancy. Let us give two realistic and illustrative examples:
\begin{itemize}
  \item A leptoquark with couplings to muon and the top quark can be used to solve the muon $g-2$ anomaly~\cite{Chakraverty:2001yg, Cheung:2001ip, Bauer:2015knc, Dorsner:2019itg}, but its contribution to $\mu$--$e$ scattering arises at one-loop level, with the leptoquark and SM fermions running in the loop, if the leptoquark couples both to electron and muon. Thus it is expected that the MUonE sensitivity to leptoquarks  is highly suppressed by the leptoquark mass and the loop factor.
  \item Models with heavy SM-singlet fermions, such as right-handed neutrinos $\nu_R$, also provide a good framework for the muon $g-2$ anomaly~\cite{Ma:2001mr, Xing:2001qn, Dicus:2001ph, Agrawal:2014ufa}. The $\mu$--$e$ scattering receives contributions from the $\nu_R - W$ loop, with $\nu_R$ coupling to both electron and muon. However, the MUonE sensitivity is expected to be highly suppressed by the loop factor, as well as by heavy-light neutrino mixings which are strongly constrained by current data (see e.g.~Ref.~\cite{Bolton:2019pcu}), apart from the $\nu_R$ mass if it is heavy.
\end{itemize}
Other such examples, like the dark photon, supersymmetric models and an $L_\mu - L_\tau$ model in the context of MUonE experiment, have been discussed in Ref.~\cite{Schubert:2019nwm}.

In light of the above considerations, we find that the most promising NP models for the MUonE experiment are light neutral scalar ($S$) and vector ($Z'$) mediators that couple to electrons and muons in either flavor-universal or non-universal way. In this paper, we show that in addition to solving (or softening) the muon $g-2$ anomaly, the existence of these particles modifies the kinematic distribution ${\rm d}\sigma   / {\rm d}T$ of $\mu$--$e$ scattering in a twofold way. First, their virtual effect is to induce new Feynman diagrams (see Figs.\ \ref{fig:feyn_tree_Zp}, \ref{fig:feyn-loop} and \ref{fig:feyn_tree_S}) that interfere with the SM contributions
(see Fig.~\ref{fig:feyn}). Second, their direct effect is to induce  new processes such as $\mu e \to \mu e S, \ \mu e Z'$, with the particles $S$ and $Z'$ decaying invisibly. If these light particles are directly produced in $\mu$--$e$ scattering, the angular distributions of outgoing electron and muon will be dramatically different from that in the $2\to 2$ process $\mu e \to \mu e$ in the SM, unless the $S$ and $Z'$ bosons are very light compared to the center-of-mass energy. In the latter case,  they tend to be very soft and it is likely that the $\mu e S$ and $\mu e Z'$-type events will be vetoed as backgrounds; therefore in this paper we will skip these direct processes and consider only the {\it virtual} effect of these light particles on MUonE.


We study the sensitivities of MUonE for these simple scenarios of light $S$ and $Z'$ bosons, and compare them with conservative existing constraints, such as those from the electron and muon $g-2$~\cite{Tanabashi:2018oca}, the searches of dark photons at BaBar~\cite{Lees:2014xha, TheBABAR:2016rlg, Lees:2017lec} and the beam-dump experiment NA64~\cite{Banerjee:2016tad, NA64:2019imj}. By ``conservative'' we mean that we do not consider ultraviolet (UV)-complete gauge-invariant scenarios that would require our new scalar and vector particles to couple also to neutrinos or even quarks, which would typically lead to stronger limits. Even in this simplistic approach, we  demonstrate that MUonE is essentially invulnerable to these  NP effects responsible for the muon $g-2$ anomaly.
Only for the case of a flavor-conserving $Z'$ boson coupling to both electrons and muons, the MUonE sensitivity is comparable to current bounds. However, we demonstrate that even this would not influence the determination of $a_\mu^{\rm had}$ at MUonE.


The rest of the paper is organized as follows:
In Section~\ref{sec:basic}, we introduce the notations and formulae
to be used in studying $\mu$--$e$ scattering. A generic discussion of NP affecting the MUonE measurement is presented in Section~\ref{sec:general_discuss}. A model-independent analysis of $Z'$ boson with tree-level couplings is detailed in Section~\ref{sec:vector}, including the scenarios of $Z'$ boson with flavor-conserving coupling in Section~\ref{sub:Flavor-conserving} and $Z'$ boson with flavor-violating couplings in Section~\ref{sub:Flavor-changing}. A model inspired by the popular $L_\mu - L_\tau$ scenario follows in Section~\ref{sec:loop}, where the $Z'$ boson couples to electrons at one-loop level.
The scalar mediators are then considered in Section~\ref{sec:scalar}, including a flavor-conserving scalar $S$ in Section~\ref{sec:scalar1} 
and a flavor-violating one in Section~\ref{sec:scalar2}.
We summarize and conclude in Section~\ref{sec:Conclusion}.



\section{MUonE Basics}\label{sec:2}

In this section we briefly outline the basics for deriving the sensitivity of the MUonE experiment to potential NP beyond the SM. We will first review the leading SM contribution to the elastic $\mu$--$e$ scattering process, and then discuss how to use the MUonE data to set limits on NP contributions. 

\subsection{Elastic $\mu$--$e$ scattering in the SM}
\label{sec:basic}


Consider a muon of energy $E_{\mu}$ scattering off an electron at rest and denote the angles of the outgoing electron and muon with respect to the incoming muon as $\theta_{e}$ and $\theta_{\mu}$, respectively.  In elastic $\mu$--$e$ scattering, there should be a distinct correlation between $\theta_{e}$ and $\theta_{\mu}$, which is crucial in selecting elastic $\mu$--$e$ scattering events from the background in the MUonE experiment. It is straightforward to obtain the angular correlation, given by
\begin{eqnarray}
\label{eq:k-3}
\cos\theta_{e} & \ = \ & \frac{E_{\mu}T+m_{e}T}{\sqrt{E_{\mu}^{2}-m_{\mu}^{2}}\ \sqrt{T(2m_{e}+T)}} \,, \\
\tan\theta_{\mu} & \ = \ &
\frac{\sqrt{T\left(2m_{e}+T\right)}\sin\theta_{e}}
{\sqrt{E_{\mu}^{2}-m_{\mu}^{2}}-\cos\theta_{e}\sqrt{T\left(2m_{e}+T\right)}} \,,
\label{eq:k-9}
\end{eqnarray}
where $T$ is the electron recoil energy, which is related to the squared momentum transfer $q^{2}$ via
\begin{equation}
q^{2} \ = \ -2m_{e}T
\ = \ \frac{x^{2}m_{\mu}^{2}}{1-x} \,,
\label{eq:qTx}
\end{equation}
with the Feynman parameter
\begin{equation}
x \ = \ \frac{\sqrt{m_{e}^{2}T^{2}+2m_{e}m_{\mu}^{2}T}-m_{e}T}{m_{\mu}^{2}} \,.
\label{eq:xT}
\end{equation}
The $\theta_{e}(T)$--$\theta_{\mu}(T)$ correlation given in Eqs.~(\ref{eq:k-3}) and (\ref{eq:k-9}) implies that given the value of $\theta_{e}$, one can determine the corresponding values of $T$ and $\theta_{\mu}$,  or vice versa (see Fig.~3 in Ref.~\cite{Abbiendi:2016xup}).


%
%

\begin{figure}
\centering
\includegraphics[width=0.25\textwidth]{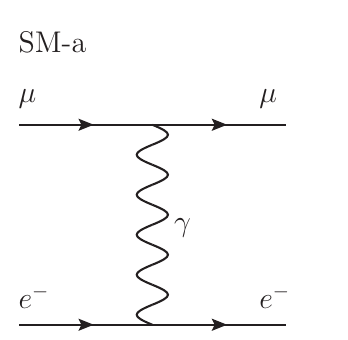}
\includegraphics[width=0.25\textwidth]{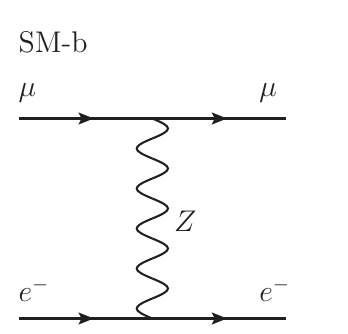}
\includegraphics[width=0.25\textwidth]{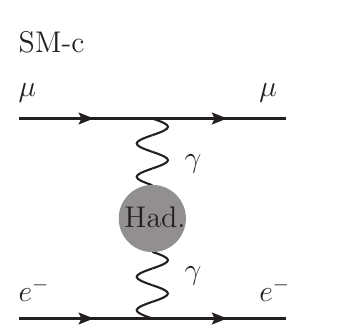}
\caption{\label{fig:feyn} Feynman diagrams for the SM processes that contribute
to $\mu$--$e$ scattering in MUonE. Compared to the dominant contribution from the SM-a diagram, the
relative sizes of the SM-b and SM-c contributions (including interference)
are about $0.001\%$ and $0.2\%$ respectively at the maximum of $q^{2}$ in MUonE.}
\end{figure}

In the MUonE experiment, with a muon beam energy of 150 GeV,  the squared center-of-mass energy is
\begin{eqnarray}
s \ = \ 2E_{\mu}m_{e}+m_{\mu}^{2}+m_{e}^{2} \ \approx \ (406\ {\rm MeV})^{2} \, ,
\end{eqnarray}
and the dominant contribution to elastic $\mu$--$e$
scattering comes from the $t$-channel QED process (see diagram SM-a
in Fig.~\ref{fig:feyn}),  which has the following differential
cross section:
\begin{equation}
\frac{{\rm d}\sigma_0}{{\rm d}T} \ = \ \frac{\pi\alpha^{2}}{(E_{\mu}^{2}-m_\mu^2)m_{e}^{2}T^{2}}\left[2E_{\mu}m_{e}\left(E_{\mu}-T\right)-T\left(m_{e}^{2}+m_{\mu}^{2}-m_{e}T\right)\right]\,,
\label{eq:SM-a}
\end{equation}
where $\alpha = {e^{2}}/{4\pi} \simeq {1}/{137}$ is the fine-structure constant. Due to the high-precision measurement in MUonE, many sub-dominant contributions
such as the neutral-current interaction (see diagram SM-b in Fig.~\ref{fig:feyn}) and the hadronic vacuum polarization
(see diagram SM-c in Fig.~\ref{fig:feyn}) are also relevant. However, for sensitivity studies, we can selectively ignore some subdominant contributions and focus on NP contributions. For this reason, we will not include the contribution of diagram SM-b or the loop-level QED corrections. The hadronic contribution (SM-c in Fig.~\ref{fig:feyn}) can be included by~\cite{Abbiendi:2016xup}
\begin{equation}
\frac{{\rm d}\sigma}{{\rm d}T} \ = \ \frac{{\rm d}\sigma_{0}}{{\rm d}T}\left|\frac{1}{1-\Delta\alpha_{{\rm had}}(q^{2})}\right|^{2} \,,
\label{eq:xsec-had}
\end{equation}
where $\Delta\alpha_{{\rm had}}(q^{2})$ denotes the running of $\alpha$
due to the hadronic contributions, which can be calculated using perturbative QCD and time-like hadro-production data (see Fig.~2 of Ref.~\cite{Abbiendi:2016xup}).
Eq.~(\ref{eq:xsec-had}) enables MUonE to experimentally measure $\Delta\alpha_{{\rm had}}(q^{2})$,
which then can be used to determine the hadronic contribution to muon
$g-2$ via the following formula \cite{Lautrup:1971jf,Calame:2015fva,Abbiendi:2016xup}:
\begin{equation}
a_{\mu}^{{\rm had}} \ = \ \frac{\alpha}{\pi}\int_{0}^{1}dx(1-x)\Delta\alpha_{{\rm had}}\left[q^{2}(x)\right].
\label{eq:a_HLO}
\end{equation}


\begin{figure}
  \centering
  \includegraphics[width=0.48\textwidth]{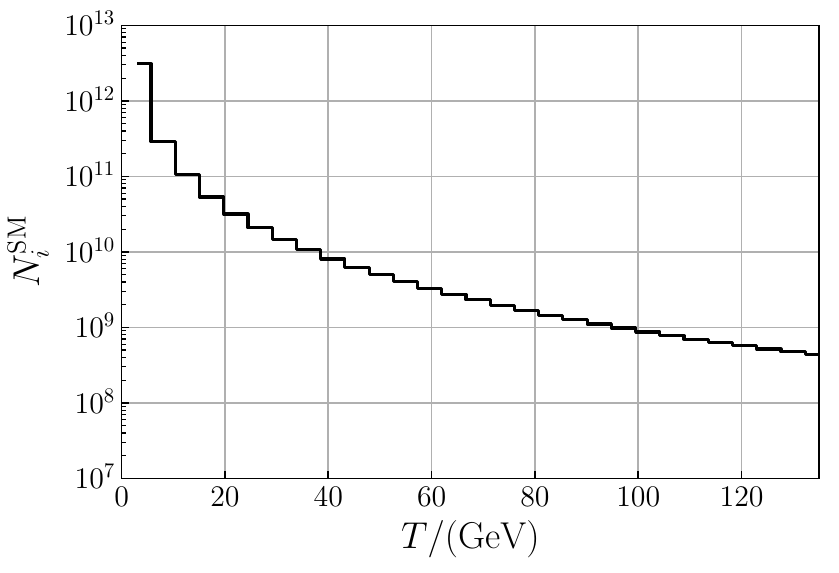}
  \includegraphics[width=0.48\textwidth]{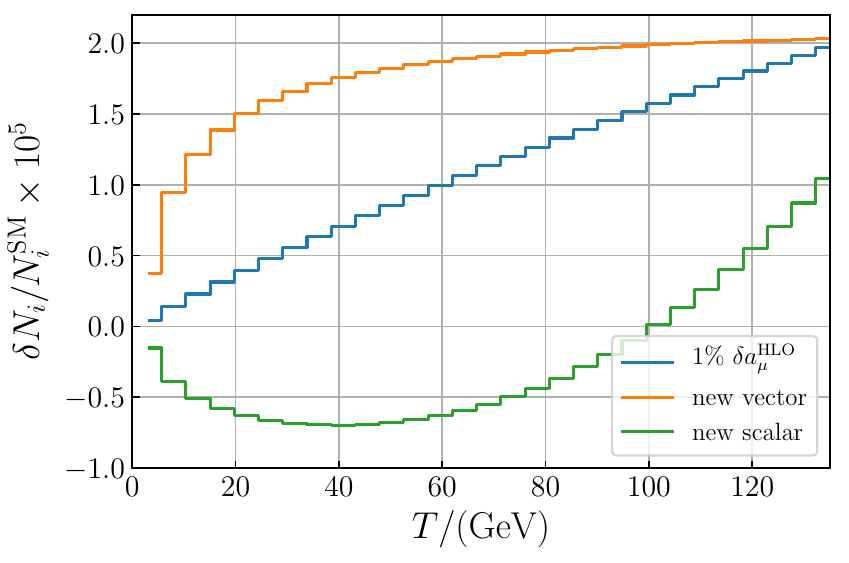}
  \caption{\label{fig:signal}
  {\it Left}: Event distribution as function of electron recoil energy $T$ in the MUonE experiment, computed at the leading-order in the SM. {\it Right}: The event excess/deficit ratios caused by 1\% larger $a_{\mu}^{{\rm had}}$ (blue), a $t$-channel $Z'$ boson (cf.~Section~\ref{sub:Flavor-conserving}) with mass $m_{Z'} = 100$ MeV and couplings $g_{Z'}^{e,\,\mu} = 10^{-3}$ (orange), and a $t$-channel $S$ (cf.~Section~\ref{sec:scalar1}) with mass $m_S = 100$ MeV and couplings $y_{ee,\,\mu\mu} = 2.4 \times 10^{-2}$ (green). }
\end{figure}

We now turn to the calculation of event rates in MUonE.
By binning the recoil spectrum ($T\rightarrow T_1, T_2, \cdots, T_i, \cdots$), the event rate in each bin can be computed by
\begin{eqnarray}
N_{i} & \ = \ & \Delta t\,N_{e}\int_{T_{i}}^{T_{i}+\Delta T}{\rm d}T\int {\rm d}E_{\mu}\Phi(E_{\mu})\times\frac{{\rm d}\sigma}{{\rm d}T}\left(T,\thinspace E_{\mu}\right)\times\Theta\left(T,\thinspace E_{\mu}\right).\label{eq:N_compute}
\end{eqnarray}
Here $N_{i}$ is event number in the $i$-th recoil energy bin ($T_{i}<T<T_{i}+\Delta T$), $\Delta t$ is the exposure time, $N_{e}$ is the number of electrons in the target, $\Phi(E_{\mu})$ is the muon flux, and the Heaviside $\Theta$ function is defined as
\begin{equation}
\Theta\left(T,\thinspace E_{\mu}\right) \ \equiv \ \begin{cases}
1 & {\rm for}\ 0<T<T_{\max}(E_{\mu})\\
0 & {\rm otherwise}
\end{cases},\label{eq:theta}
\end{equation}
where $T_{\max}$ is the maximal recoil energy in the MUonE experiment, purely determined by kinematics:
\begin{equation}
T_{\max} \ = \ \frac{2m_{e}\left(E_{\mu}^{2}-m_{\mu}^{2}\right)}{2E_{\mu}m_{e}+m_{e}^{2}+m_{\mu}^{2}} \ \approx \ 140\ {\rm GeV}.\label{eq:Tmax}
\end{equation}
In practice, due to the monochromatic muon beam in MUonE, we can consider
$\Phi(E_{\mu})$ as a delta function, so that Eq.~(\ref{eq:N_compute})
can be further simplified to
\begin{equation}
N_{i} \ = \ L\int_{T_{i}}^{T_{i}+\Delta T}\frac{{\rm d}\sigma}{{\rm d}T}\left(T,\thinspace E_{\mu}\right){\rm d}T\,,\label{eq:N_L}
\end{equation}
where $L=\Delta t\,N_{e}\int {\rm d} E_{\mu}\Phi(E_{\mu})$ is the integrated
luminosity. In the MUonE experiment with two years of data-taking,
the integrated luminosity is expected to reach $L=1.5\times10^{7}\ {\rm nb}^{-1}$
\cite{Abbiendi:2016xup}.  With the SM total cross section $\sigma_0=245\ {\rm \mu b}$, one can reach a large number of total events:
\begin{equation}
N_{{\rm total}} \ = \ L\,\sigma_0 \ \approx \ 3.7\times10^{12}\,.\label{eq:N_tot}
\end{equation}
The event distribution with respect to the electron recoil energy $T$ in the MUonE experiment is shown in the left panel of Fig.~\ref{fig:signal} (see also Fig.~4 in Ref.~\cite{Abbiendi:2016xup}). In the right panel of Fig.~\ref{fig:signal}, the blue line illustrates the effect on the recoil spectrum of an $a_{\mu}^{{\rm had}}$ value enlarged by 1\% with respect to the central value given in Eq.\ (\ref{eq:a-had}).
For comparison, the effect of two possible NP contributions is shown, one with a $t$-channel light gauge boson $Z'$ with mass of 100 MeV and a coupling of $10^{-3}$ (see Section~\ref{sub:Flavor-conserving}), and the other one with a light neutral scalar mediator $S$ with mass of 100 MeV and a coupling of $2.4 \times 10^{-2}$ (see Section~\ref{sec:scalar1}). Here $N_{i}^{{\rm SM}}$ and $N_i$ are respectively the expected event numbers in the $i$-th bin in the SM and in the presence of NP (or anomalous hadronic vacuum polarization).

In this work we adopt the following simple $\chi^{2}$-function to estimate the sensitivities of MUonE to the anomalous hadronic vacuum polarization (and to NP):
\begin{equation}
\chi^{2} \ = \
\sum_{i}\frac{(N_{i}-N_{i}^{{\rm SM}})^{2}}{\sigma_{{\rm stat},i}^{2}+\sigma_{{\rm sys},i}^{2}} \, ,
\label{eq:chi2}
\end{equation}
where $\sigma_{{\rm stat},i}=\sqrt{N_{i}}$ are the statistical uncertainties, and $\sigma_{{\rm sys},i}=10^{-5}N_{i}$ are the systematic uncertainties at the level of 10 ppm~\cite{Abbiendi:2016xup}.


\subsection{NP effects on MUonE experiment}
\label{sec:general_discuss}

NP explanations for the muon $g-2$ anomaly have been comprehensively reviewed e.g.\ in Refs.~\cite{Jegerlehner:2009ry,Lindner:2016bgg}. In general, one needs to introduce either a (light) $Z'$ gauge boson or a (light) neutral scalar $S$ that interacts with muons
to address the anomaly. The contribution of a $Z'$ gauge boson to the anomalous magnetic moment of a charged lepton ($\ell=e,\ \mu,\ \tau$)
is in general given by~\cite{Leveille:1977rc}:
\begin{equation}
\Delta a_{\ell} \ = \
\frac{g_{Z'}^{2}\epsilon_{Z'}^{2}}{8\pi^{2}}\int_{0}^{1} {\rm d} x
\frac{x^{2}(1-\epsilon_{\ell'})^{2}(1-x+\epsilon_{\ell'})
\epsilon_{Z'}^{2}+2x(1-x)(x+2\epsilon_{\ell'}-2)}{(1-x)
\left(1-x\epsilon_{Z'}^{2}\right)+x\epsilon_{\ell'}^{2}
\epsilon_{Z'}^{2}} \,,
\label{eq:g-2_v}
\end{equation}
where $g_{Z'}$ is the gauge coupling of $Z'$ to the charged leptons, $\epsilon_{Z'}\equiv m_{\ell}/m_{Z'}$, $\epsilon_{\ell'}\equiv m_{\ell'}/m_{\ell}$ (with $m_{Z'}$ and $m_\ell$ being the $Z'$ and charged lepton masses respectively), and $\ell'$ is the lepton running in the loop of the corresponding $g-2$ diagram. Similarly, the contribution of a neutral scalar $S$ is formulated as~\cite{Lindner:2016bgg}
\begin{eqnarray}
\Delta a_{\ell}\ = \ \frac{|y_{\ell\ell^{\prime}}|^{2}}{8\pi^{2}}
\frac{m_{\ell}^{2}}{m_{S}^{2}}\int_{0}^{1}{\rm d}x\frac{x^{2}(1-x \pm \epsilon_{\ell'})}
{(1-x)(1-x\epsilon_{S}^{2})+x\epsilon_{\ell'}^{2}\epsilon_{S}^{2}}\,,
\label{eq:g-2_s}
\end{eqnarray}
where $y_{\ell\ell^{\prime}}$ is the Yukawa coupling of $S$ to
the leptons, the plus (minus) sign in the numerator is for the case of CP-even (CP-odd) neutral scalar, and $\epsilon_{S}\equiv m_{\ell}/m_{S}$ with $m_S$ the neutral scalar mass. Note that for both gauge and scalar mediators the charged lepton flavors $\ell$ and $\ell'$ can be the same or different.
Eqs.~(\ref{eq:g-2_v}) and (\ref{eq:g-2_s}) are obtained by simply evaluating the $F_{2}$
form factor of the triangle diagrams with $Z'$ or $S$ and $\ell'$
running in the loop. Since the $F_{2}$ form factor is always finite
in triangle diagrams, the results are free from UV divergence. However,
one should note that here we are discussing a generic $Z'$ or $S$
without referring to the complete models, which may contain other
new particles that also contribute to $(g-2)_{\ell}$.

According to the current theoretical and experimental values of the muon $g-2$ \cite{Tanabashi:2018oca}, we are most interested in  new vector or scalar interactions that can produce, or contribute to, the theoretical-experimental discrepancy given by Eq.~\eqref{eq:a_mu_val},
while satisfying all other relevant laboratory, astrophysical and cosmological constraints, in particular those from electron $g-2$, BaBar and NA64.
%
%
%
In the presence of new gauge or scalar interactions (that can potentially address the muon $g-2$ anomaly), we want to know whether or to what extent these NP interactions can affect the MUonE measurement. In other words, if the MUonE measurement of the hadronic polarization turns out to be inconsistent with the SM expectation, we need to examine whether it is due to the same NP that causes the muon $g-2$ anomaly, or simply due to an inaccurate QCD calculation. We summarize below the various scalar or vector boson scenarios, and discuss their relevance to the MUonE experiment. Before going into detail, we stress again that we will take a conservative approach and do not consider full UV-complete gauge-invariant scenarios, in which the bosons would necessarily couple to neutrinos or quarks with the same strength as to charged leptons. Those would be subject to several other stronger limits than the ones we consider here, and thus even strengthen our conclusions on the relevance of NP for the MUonE sensitivity. Alternatively, it might be also possible that the breaking of the symmetry associated to the $Z'$ boson or $S$ mass takes place below the electroweak scale, and the couplings of $Z'$ or $S$ to other SM particles are much smaller than those to the charged leptons.

For the case of light $Z'$ boson, we consider the following three possibilities:
\begin{enumerate}[V(i)]
\item {\it Flavor-universal gauge interactions in the $t$-channel}: These interactions can in general be formulated as
  \begin{equation}
  {\cal L} \ \supset \ g_{Z'} Z_\mu^\prime
  \left( \bar{e} \gamma^\mu e + \bar{\mu} \gamma^\mu \mu \right),
  \label{eq:z}
  \end{equation}
  which is the most widely considered scenario of vector mediators. The flavor-universal interactions in Eq.~(\ref{eq:z}) cause a $t$-channel process in MUonE, which will be analyzed in detail in Section~\ref{sub:Flavor-conserving}. The case includes the $U(1)_{B-L}$ gauge extension of the SM
  and applies also to dark photon models~\cite{Fayet:2007ua, Pospelov:2008zw, Alexander:2016aln}. In the latter case the dark photon interacts with charged leptons via kinetic mixing with the SM photon, and one only needs to replace the coupling $g_{Z'}$ in Eq.~(\ref{eq:z}) by $\epsilon e$, with $\epsilon$ being the kinetic mixing parameter.

  \item {\it Flavor-non-universal gauge interactions in the $t$-channel}: In general the interaction can be written as
      \begin{equation}
      {\cal L} \ \supset \ Z_\mu^\prime
      \left( g_{Z'}^{e} \bar{e} \gamma^\mu e + g_{Z'}^{\mu} \bar{\mu} \gamma^\mu \mu \right),
      \label{eq:z-FNU}
      \end{equation}
      where $g_{Z'}^{e}\neq g_{Z'}^{\mu}$.  Eq.~(\ref{eq:z-FNU}) serves as the most general model-independent formula to study a $t$-channel $Z'$ mediator; Eq.~\eqref{eq:z} for the case V(i) can simply be obtained by setting $g_{Z'}^{e}=g_{Z'}^{\mu}\equiv g_{Z'}$. The MUonE sensitivities for cases V(i) and  V(ii) are presented in Fig.~\ref{fig:constraints_Z_t}.
      This scenario typically arises from models with gauged lepton numbers.
      The anomaly-free possibilities of gauged lepton numbers include $L_{e}-L_{\mu}$, $L_{\mu}-L_{\tau}$, $L_{e}-L_{\tau}$, and any of their linear combinations. More generally, gauged $Q_e L_{e}+Q_\mu L_{\mu}+Q_\tau L_{\tau}$ with $Q_e+Q_\mu+Q_\tau=0$ is always anomaly free, and this causes $g_{Z'}^{e}/g_{Z'}^{\mu}=Q_e/Q_\mu$. Sometimes the case $L_e - L_\mu$ is discussed~\cite{Asai:2018ocx}.
      There is however another  interesting model in this category, namely the $L_{\mu}-L_{\tau}$ model \cite{He:1990pn,Foot:1990mn,He:1991qd}. In this model, which provides a reasonable first-order approximation to lepton mixing~\cite{Ma:2001md,Heeck:2011wj}, $g_{Z'}^{e}=0$ at tree-level, and non-vanishing $g_{Z'}^e$ can arise at one-loop level. We evaluate this coupling, which to our knowledge has not been done before.
      It is remarkable that the $L_{\mu}-L_{\tau}$ model still provides a viable explanation of the muon $g-2$ anomaly even after all experimental constraints are imposed~\cite{Gninenko:2018tlp, Garani:2019fpa} (see Fig.~\ref{fig:constraints-mutau}).

  \item {\it Flavor-changing gauge interactions in the $s$-channel}: The couplings of $Z'$ boson to the SM leptons can not only be flavor-non-universal but also flavor-violating~\cite{Heeck:2016xkh, Altmannshofer:2016brv}. For instance, we can have the couplings
      \begin{equation}
      {\cal L} \ \supset \ g_{Z'}^{e\mu} Z_\mu^\prime
      \bar{e} \gamma^\mu \mu ~+~ {\rm H.c.},
      \label{eq:z-FC}
      \end{equation}
      which can induce elastic $\mu$--$e$ scattering in the $s$-channel. The flavor-violating coupling $g_{Z'}^{e\mu}$ can be potentially constrained by the MUonE data, as shown in Fig.~\ref{fig:constraints_Z_s}. For simplicity, we will consider in this paper only the coupling $g_{Z'}^{e\mu}$ but not other possibilities of lepton flavor combinations such as $g_{Z'}^{\mu\tau}$ as these are not directly relevant to the $\mu$--$e$ scattering and the MUonE experiment.

\end{enumerate}


As for the light neutral scalar mediator $S$, we consider the following two possibilities:
\begin{enumerate}[S(i)]
  \item {\it Flavor-conserving Yukawa interactions in the $t$-channel}: The most general flavor-conserving scalar couplings to electron and muon can be written as
      \begin{equation}
      {\cal L} \ \supset \ S
      \left( y_{ee} \bar{e}e + y_{\mu\mu} \bar{\mu}\mu \right),
      \label{eq:s-FNU}
      \end{equation}
      where $y_{ee}$ and $y_{\mu\mu}$ are respectively the new Yukawa couplings for electron and muon, which in general are different. Here for the sake of concreteness we have assumed the scalar $S$ to be CP-even.\footnote{We have also checked the electron and muon $g-2$ limits for a light CP-odd scalar $A$. The resultant $g-2$ limits on $A$ are slightly weaker than those for $S$, but our conclusions are not affected.}  The couplings in Eq.~(\ref{eq:s-FNU}) contribute to elastic $\mu$--$e$ scattering in the $t$-channel. The MUonE sensitivities can be found in Fig.~\ref{fig:scalar_r}. As shown explicitly in Eq.~(\ref{eq:z-2}) and Eq.~(\ref{eqn:cs:S1}) below, the major difference between the scalar and gauge boson mediated $t$-channel $\mu$--$e$ scattering process is that the interference between the SM and new physic contributions is comparatively suppressed by a factor of $m_{\mu}/E_{\mu}$ for the scalar case. This can be easily understood from the chirality-flipping nature of scalar interactions.  As a consequence of the suppressed interference effect, the MUonE experiment is less sensitive to a $t$-channel scalar mediator than a $t$-channel vector mediator.

  \item {\it Flavor-changing Yukawa interactions in the $s$-channel}: Similar to Eq.~(\ref{eq:z-FC}), we also consider the following flavor-changing scalar interactions:
    \begin{equation}
    {\cal L} \ \supset \
    y_{e\mu} S \bar{e} \mu ~+~ {\rm H.c.}\,
    \label{eq:s-FC}
    \end{equation}
    As for the gauge coupling case, this coupling leads to an $s$-channel $\mu$--$e$ scattering process and might get constrained by the MUonE events, as shown in Fig.~\ref{fig:scalar2}. As in the gauge coupling case, we will not consider other flavor combinations such as $y_{\mu\tau}$ as these are not directly relevant to the $\mu$--$e$ scattering and the MUonE experiment.

\end{enumerate}
For completeness, we would also like to mention the possibility of a doubly-charged scalar that has a flavor-violating coupling to electrons and muons, which would generate elastic $\mu$--$e$ scattering in the $u$-channel. Its mass is however constrained to be above a few hundred GeV, in particular from LHC searches~\cite{Dev:2018kpa}, and its contribution to the magnetic moment of the muon would always be negative, i.e.\ goes in the wrong direction~\cite{Dev:2018upe}.





\section{Vector mediators}
\label{sec:vector}

\begin{figure}[!t]
  \centering
  \includegraphics[height=0.25\textwidth]{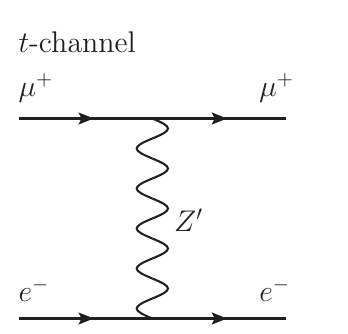}\hspace{1cm}
  \includegraphics[height=0.25\textwidth]{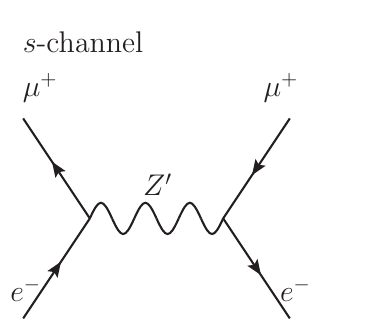}
  \caption{Feynman diagrams of $\mu$--$e$ scattering mediated by a flavor-conserving $Z'$ boson in the $t$-channel (left) or a flavor-changing $Z'$ boson in the $s$-channel (right). }
\label{fig:feyn_tree_Zp}
\end{figure}

In this section, we discuss the MUonE sensitivity to a generic light vector mediator $Z'$ in both $t$ and $s$ channels and with either flavor-conserving or violating couplings to electrons and muons at tree-level.

\subsection{Flavor-conserving couplings in the $t$-channel}
\label{sub:Flavor-conserving}

Both the cases V(i) and V(ii) for the $Z'$ gauge boson have flavor-conserving interactions, which contribute to elastic $\mu$--$e$ scattering via the $t$-channel process shown in the left diagram in Fig.~\ref{fig:feyn_tree_Zp}. Let us start with the case V(ii) with flavor-dependent couplings of a $Z'$ boson; the case V(i) with flavor-universal couplings can be simply obtained by setting $g_{Z'}^{e}=g_{Z'}^{\mu}$. In the presence of the $t$-channel $Z'$-mediated process, it is straightforward to calculate the cross section for elastic $\mu$--$e$ scattering, which reads:
\begin{equation}
\frac{d\sigma}{dT} \ = \
\left.\frac{d\sigma}{dT}\right|_{{\rm SM}}
\times\left[ 1+ \frac{g_{Z'}^{e}g_{Z'}^{\mu}m_{e}T(2m_{e}T
+m_{Z'}^{2})}{\pi\alpha\left(m_{Z'}^{2} +2m_{e}T\right)^{2}}
+\frac{\left(g_{Z'}^{e}g_{Z'}^{\mu}m_{e}T\right)^{2}}
{4\pi^{2}\alpha^{2}\left(m_{Z'}^{2}+2m_{e}T\right)^{2}}\right].
\label{eq:z-2}
\end{equation}
This expression includes the SM cross section (first term in the bracket) proportional to $e^{4}$, the $Z'$-mediated cross section (last term) proportional to $(g_{Z'}^{e} g_{Z'}^{\mu})^2$, and the interference between the SM and $Z'$-mediated process (middle term) proportional to $(g_{Z'}^{e} g_{Z'}^{\mu}) e^2$. When the $g_{Z'}$ couplings are small, i.e., $g_{Z'}^{e,\,\mu} \ll e$, the interference term is more important than the $(g_{Z'}^{e} g_{Z'}^{\mu})^2$ term and one can expect that the MUonE sensitivity is approximately proportional to $g_{Z'}^{e} g_{Z'}^{\mu}$.

With the cross section given by Eq.~(\ref{eq:z-2}) and using the $\chi^{2}$-function introduced in Eq.~\eqref{eq:chi2}, it is straightforward to compute the sensitivity of MUonE on the coupling product $g_{Z'}^{e}g_{Z'}^{\mu}$. For the cases V(i) and V(ii) with flavor-conserving couplings, the electron and muon  $g-2$ depend only on the couplings $g_{Z'}^{e}$ and $g_{Z'}^{\mu}$ respectively, and the $\mu$--$e$ scattering needs both couplings. To compare the MUonE sensitivity with current bounds and other future prospects, we introduce a ratio $r$ of the electron and muon gauge couplings which is defined as
\begin{equation}
r_V \ \equiv \
\sqrt{ \frac{g_{Z'}^{e}}{g_{Z'}^{\mu}} } \,.
\label{eq:r_def}
\end{equation}
With this definition of $r_V$, the differential cross section ${\rm d}\sigma/{\rm d}T \propto g_{Z'}^e g_{Z'}^\mu \propto r_V^2$, and thus the MUonE sensitivities scale simply as $r_V$.
The cases V(i) and V(ii) correspond respectively to the values of $r_V = 1$ and $r_V \neq 1$. The MUonE prospects of $m_{Z'}$ and $g_{Z'}^\mu$ and the current constraints for the case V(i) with  $g_{Z'}^e = g_{Z'}^\mu$ (and equivalently $r_V = 1$) are shown in the upper panels of Fig.~\ref{fig:constraints_Z_t}. For the purpose of concreteness, we take three example values $r_V = \{ 0.8,\ 0.6,\ 0.4\}$
for the case V(ii). The corresponding MUonE prospects, as well as current limits from other searches, are shown respectively in the upper middle, lower middle and lower panels in Fig.~\ref{fig:constraints_Z_t}. For $r_V>1$, the electron $g-2$, BaBar and NA64 bounds (see below) will move further down, leaving no space for MUonE to see a signal.
It should be noted that if the $Z'$ couplings to electron and muon have opposite signs such as in the $L_{e}-L_\mu$ model, the value of $r_V^2$ may be {\it negative}. The MUonE sensitivity for negative $r_V^2$ is almost the same as for positive $r_V^2$ because the interference term in the cross section, which is the dominant NP contribution, is proportional to $r_V^2$. In particular $r_V^2=-1$ leads to almost the same $|\delta N_i|$ as $r_V=1$ does, but the sign of $\delta N_i$ is opposite. This also holds true for the scalar case in Section~\ref{sec:scalar}.

There are a large variety of experimental constraints on a light $Z'$ boson, including those from electron and muon $g-2$~\cite{Tanabashi:2018oca}, beam dump experiments~\cite{Bjorken:2009mm, Batell:2009di, Essig:2010gu}, big bang nucleosynthesis (BBN)~\cite{Kamada:2015era,Huang:2017egl, Kamada:2018zxi}, supernovae~\cite{Dent:2012mx,Kazanas:2014mca,Farzan:2018gtr}, non-standard neutrino interactions (NSI)~\cite{Heeck:2018nzc}, neutrino-electron scattering \cite{Bilmis:2015lja,Lindner:2018kjo,Arcadi:2019uif,Link:2019pbm}, coherent elastic neutrino-nucleus scattering (CE$\nu$NS) \cite{Dutta:2015vwa,Abdullah:2018ykz,Aguilar-Arevalo:2019zme}, the BaBar experiment~\cite{Lees:2014xha, TheBABAR:2016rlg}, etc. Some of these potential bounds depend on the $Z'$ couplings to quarks, charged leptons or neutrinos, while some of them are not relevant to the parameter space of the MUonE experiment we are considering here.  After reviewing all the possible bounds discussed in the literature, we find that the most relevant limits for the MUonE sensitivities are those from the electron and muon $g-2$, the BaBar experiment \cite{Lees:2014xha,TheBABAR:2016rlg,Lees:2017lec} and the NA64 experiment \cite{Banerjee:2016tad,NA64:2019imj}.
Here we would like to stress again that we conservatively do not impose constraints derived from the SM gauge invariance or any UV-complete model.

Using Eq.~(\ref{eq:g-2_v}), it is straightforward to evaluate the $Z'$ contribution to muon $g-2$ with $\epsilon_{\ell'}=1$. The purple shaded regions in Fig.~\ref{fig:constraints_Z_t} are excluded at  $5\sigma$ confidence level (C.L.),
and the green bands indicate the parameter space of $m_{Z'}$ and $g_{Z'}^\mu$ for the muon $g-2$ anomaly at  $1\sigma$ C.L.  Similarly, one can obtain the electron $g-2$ limits from the $2\sigma$ upper limit $\Delta a_{e}<5.0\times10^{-13}$~\cite{Tanabashi:2018oca}\footnote{As pointed out in Ref.~\cite{Davoudiasl:2018fbb}, a recent measurement of the fine structure constant~\cite{Parker:2018vye} leads to a $2.4\sigma$ discrepancy between the theoretical and experimental values of the electron $g-2$, i.e. $\Delta a_e=(-8.7\pm 3.6)\times 10^{-13}$. However, since this is opposite in sign to the muon $g-2$ anomaly [cf.~Eq.~\eqref{eq:a_mu_val}], it is difficult to address both of them in the simple NP model frameworks studied here~\cite{Gardner:2019mcl, CarcamoHernandez:2019ydc}, although suitable UV-complete extensions can do this~\cite{Davoudiasl:2018fbb, Crivellin:2018qmi, Liu:2018xkx, Han:2018znu, Endo:2019bcj, Abdullah:2019ofw, Badziak:2019gaf, Hiller:2019mou}.  Therefore, we simply use the $2\sigma$ upper limit for $\Delta a_e$ quoted in Ref.~\cite{Tanabashi:2018oca}.} on $m_{Z'}$ and $g_{Z'}^\mu$ for all the four cases of $r_V = \{ 1,\ 0.8,\ 0.6,\ 0.4\}$, which are also presented in Fig.~\ref{fig:constraints_Z_t}.


\afterpage{%
\begin{figure}
  \centering
  \vspace{-1cm}
  \includegraphics[height=18.5cm]{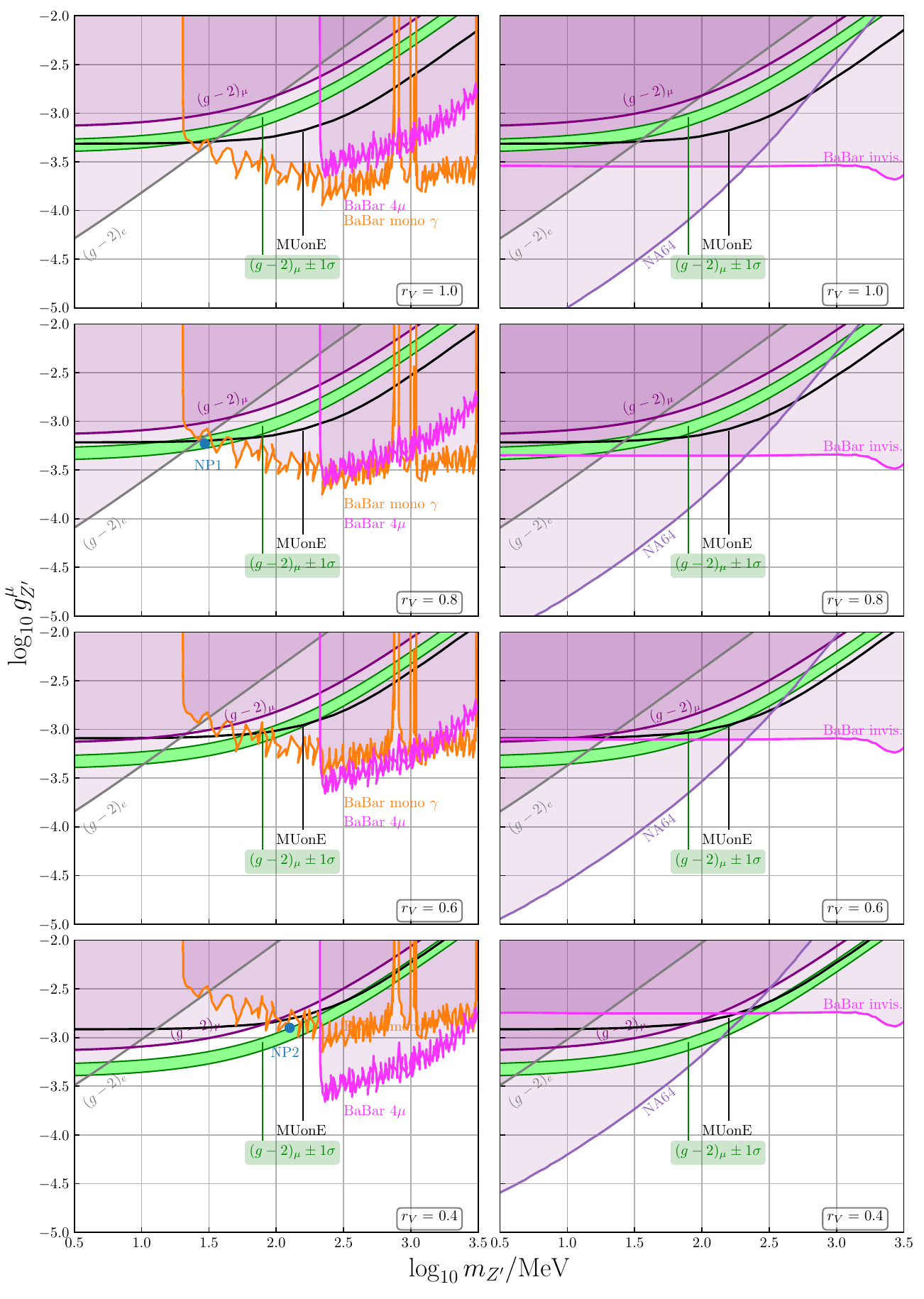}
  \caption{Sensitivities of MUonE on a $t$-channel vector mediator with the interactions given in Eq.~(\ref{eq:z-FNU}) for $r = 1$ (upper panel), 0.8 (upper middle), 0.6 (lower middle) and 0.4 (lower). We also show the limits from electron and muon $g-2$, BaBar monophoton, 4$\mu$ and invisible searches~\cite{Lees:2014xha,TheBABAR:2016rlg,Lees:2017lec}, and from NA64~\cite{Banerjee:2016tad,NA64:2019imj}. The green regions correspond to the required values of $m_{Z'}$ and $g_{Z'}$ to explain the muon $g-2$ anomaly at the $1\sigma$ C.L. Depending on whether $Z'$ dominantly decays into visible (left panel) or invisible particles (right panel), the BaBar and NA64 bounds apply differently. NP1 in the left upper middle panel and NP2 in the left lower panel are two benchmark points that are shown in Fig.~\ref{fig:a_had}; see Eq.~(\ref{eqn:NP}) for the benchmark values.  }
\label{fig:constraints_Z_t}
\end{figure}
\clearpage
}

The $Z'$ boson can be produced in the BaBar experiment via the processes
\begin{eqnarray}
\label{eqn:Z'}
e^{+}e^{-}\rightarrow\gamma Z' \quad \text{and} \quad
e^{+}e^{-}\rightarrow\mu^{+}\mu^{-}Z'\,.
\end{eqnarray}
It may further decay into electrons or muons ($Z' \to e^+ e^-,\, \mu^+ \mu^-$), or into some invisible particles
such as neutrinos or other light dark particles.
Depending on whether $Z'$ decays dominantly into visible or invisible final states, one needs to consider different BaBar bounds.
For visible decay, we take two bounds from Refs.~\cite{Lees:2014xha,TheBABAR:2016rlg}. The bound published in Ref.~\cite{Lees:2014xha} measures the $\gamma Z'$ process in Eq.~(\ref{eqn:Z'}) with a monophoton, assuming a flavor-universal coupling. Consequently, the $g^e_{Z'}$ coupling is responsible for $Z'$ production and the bound is mainly sensitive to $g^e_{Z'}$. The mono $\gamma$ limits on $m_{Z'}$ and $g_{Z'}^\mu$ are presented as the orange shaded regions in the left panels of Fig.~\ref{fig:constraints_Z_t}.
Another BaBar bound published in Ref.~\cite{TheBABAR:2016rlg} assumes that $Z'$ exclusively couples to $\mu$, which means that $Z'$ is produced via $e^{+}e^{-}\rightarrow\mu^{+}\mu^{-}Z'$, producing a $4\mu$ signal, and can thus be interpreted as a bound on $g^{\mu}_{Z'}$, as indicated by the pink shaded regions in the left panels of Fig.~\ref{fig:constraints_Z_t}.
If $Z'$ decays invisibly, then we take the bound from Ref.~\cite{Lees:2017lec}, which mainly constrains $g^e_{Z'}$, as shown by the pink shaded regions in the right panels of Fig.~\ref{fig:constraints_Z_t}.

The NA64 experiment is an electron beam dump experiment, specifically designed for the invisible decay scenarios. With its coupling to electrons, the $Z'$ boson can be produced in NA64 from bremsstrahlung off the electron beam. The invisibly decaying $Z'$ boson can be searched for by measuring the missing energy. The NA64 bounds in Refs.~\cite{Banerjee:2016tad,NA64:2019imj} exclude the dark blue shaded regions in the right panels of Fig.~\ref{fig:constraints_Z_t}.

It is clear from the two upper panels of Fig.~\ref{fig:constraints_Z_t} that, for the case V(i) with $g_{Z'}^e = g_{Z'}^\mu$ (and thus $r_V=1$), no matter whether the $Z'$ boson decays into visible or invisible particles, the MUonE sensitivity is weaker than  the limits from electron $g-2$ and BaBar data (and NA64). When the ratio $r_V$ gets smaller, the coupling $g_{Z'}^e$ will be getting smaller compared to $g_{Z'}^\mu$, thus the electron $g-2$ limits will get weaker. Let us consider separately the two cases:
\begin{itemize}
  \item If the dominant decay channel is $Z' \to e^+ e^- ,\, \mu^+ \mu^-$, when the coupling $g_{Z'}^e$ gets smaller, the $e^+ e^- \to \gamma Z'$ process at BaBar will be suppressed, and the $e^+ e^- \to 4 \mu$ process is not affected.   As shown in the left panels of Fig.~\ref{fig:constraints_Z_t},
      for $r_V = 0.4$, the MUonE prospect has been precluded by the limits from muon $g-2$ and the BaBar data, although there is a narrow window left at the edge of the BaBar limit for $r_V = 0.6$ and 0.8.

  \item For the case of invisibly decaying $Z'$ boson, as shown in the right panels of Fig.~\ref{fig:constraints_Z_t}, the NA64 limits are more stringent than those from electron $g-2$ for all four scenarios with $r_V = \{ 1, \ 0.8, \ 0.6,\ 0.4 \}$. As a result, the invisible search limits from NA64 and BaBar have precluded the MUonE sensitivity in this case. 
\end{itemize}

We should not overinterpret the small windows of opportunity in Fig.\ \ref{fig:constraints_Z_t} due to the statistical fluctuations in the BaBar limit, but they offer two interesting benchmark points, namely
\begin{equation}
\label{eqn:NP}
\begin{array}{rlll}
{\rm NP1:} & r_V \ = \ 0.8\,, \, & ~ m_{Z'} \ = \ 29.5 \, {\rm MeV} \,, & ~ g_{Z'}^\mu \ = \ 5.89 \times 10^{-4} \,,\\
{\rm NP2 (\pm):} & |r_V| \ = \ 0.4 \,, & ~ m_{Z'} \ = \ 126 \, {\rm MeV} \,, & ~ g_{Z'}^\mu \ = \ 1.26 \times 10^{-3} \,.
\end{array}
\end{equation}
The benchmark point NP1 lies in the
$1\sigma$ band of the muon $g-2$, is at the borderline of MUonE's sensitivity and BaBar's exclusion limits (see the left upper middle panel in Fig.~\ref{fig:constraints_Z_t}). The second benchmark point NP2 is slightly above the $1\sigma$ band of the muon $g-2$ (see the left lower panel).

\begin{figure}[t!]
  \centering
  \includegraphics[width=0.8\textwidth]{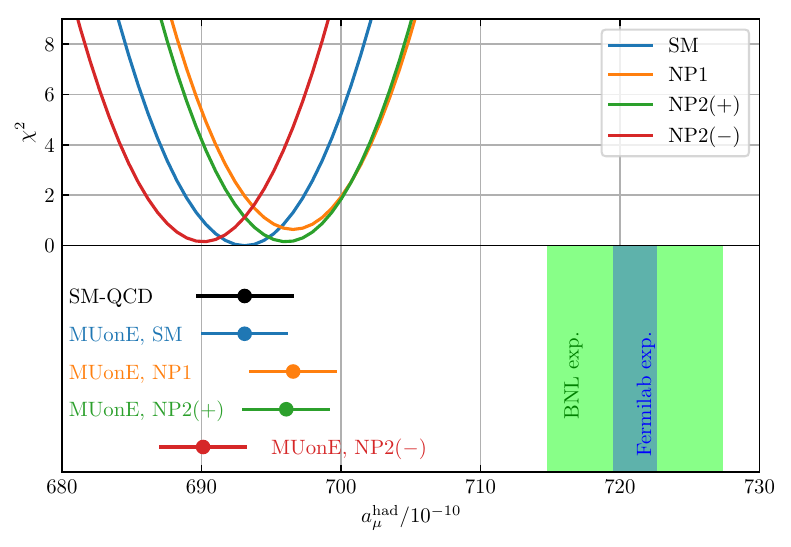}
  \caption{Impact of NP on the MUonE measurement of $a^{\rm had}_{\mu}$. Here NP1 and NP2 are two benchmark scenarios of the $t$-channel $Z'$ model marked in Fig.~\ref{fig:constraints_Z_t}, and the parameters are given in Eq.~(\ref{eqn:NP}). The upper part show the $\chi^2$-curves for the SM, NP1 and NP2$(\pm)$, while the $1\sigma$ ranges of the SM QCD prediction of $a_\mu^{\rm had}$, the MUonE sensitivity in the SM, NP1 and NP2$(\pm)$ are presented in the lower part. Also shown are the value of $a_\mu^{\rm had}$ from current BNL experimental value of $g-2$  (green)~\cite{Bennett:2006fi} and future prospect at Fermilab (blue)~\cite{Grange:2015fou}.
   }
\label{fig:a_had}
\end{figure}

To illustrate the difference of positive and negative $r_V^2$, we adopt two cases for NP2, with NP2(+) corresponding to a positive $r_V^2 = 0.4^2$ while NP2$(-)$ for $r_V^2 = - 0.4^2$, for which the $Z'$ couplings to electron and muon have opposite signs. Note that, neglecting the $(g_{Z'}^e g_{Z'}^\mu)^2$ term in Eq.~(\ref{eq:z-2}), the MUonE sensitivities of NP2$(\pm)$ are the same. The impact of NP1 and NP2$(\pm)$ on the determination of the hadronic contribution at MUonE is illustrated in Fig.\ \ref{fig:a_had}.
Using Eq.~\eqref{eq:chi2},
the $\chi^2$-distributions are shown in the upper part of the figure.
For the SM $a_\mu^{\rm had}$ we take the central value from Eq.\  (\ref{eq:a-had}).
In the lower part of Fig.\ \ref{fig:a_had} we show the $1\sigma$ range (0.6\%) of the SM value of $a_\mu^{\rm had}$ in Eq.\  (\ref{eq:a-had}), the claimed improvement to 0.3\% by MUonE and the $1\sigma$ sensitivities  we obtain for the benchmark points NP1 and NP2$(\pm)$. It is very clear in Fig.~\ref{fig:a_had} that, as a result of the different $Z'$ couplings to electron and muon, NP2(+) and NP2$(-)$ move in opposite directions with respect to the SM prediction of $a_\mu^{\rm had}$.
However, both the benchmark points we are considering, which are currently marginally allowed by other searches, do not have a significant effect in the determination of $a_\mu^{\rm had}$ at MUonE, as all $1\sigma$ ranges in Fig.~\ref{fig:a_had} overlap (partially) with each other. As the MUonE sensitivity for all other scenarios below have been precluded by the current limits (see Figs.~\ref{fig:constraints_Z_s}, \ref{fig:constraints-mutau}, \ref{fig:scalar_r} and \ref{fig:scalar2}), we do not include benchmark points for any other scenarios in Fig.~\ref{fig:a_had}.



\subsection{Flavor-changing couplings in the $s$-channel}
\label{sub:Flavor-changing}

As mentioned in Section~\ref{sec:general_discuss}, there is only one flavor-changing gauge coupling $g_{Z'}^{e\mu}$ that is relevant to the MUonE experiment [see Eq.~(\ref{eq:z-FC})]. It contributes to elastic $\mu$--$e$ scattering via the $s$-channel diagram shown in Fig.~\ref{fig:feyn_tree_Zp}. Note that in the absence of $Z'$--$e$--$e$ and $Z'$--$\mu$--$\mu$ couplings the flavor-changing coupling does not necessarily lead to charged lepton flavor violating (LFV) processes such as $\mu \to e\gamma$ and $\mu \to eee$. Vector bosons that couple to different lepton flavors may originate from gauged flavor symmetry models~\cite{Heeck:2016xkh}.
Although they are less common in beyond SM theories, an $s$-channel resonance in the MUonE experiment can be potentially interesting.

\begin{figure}
  \centering
  \includegraphics[width=0.65\textwidth]{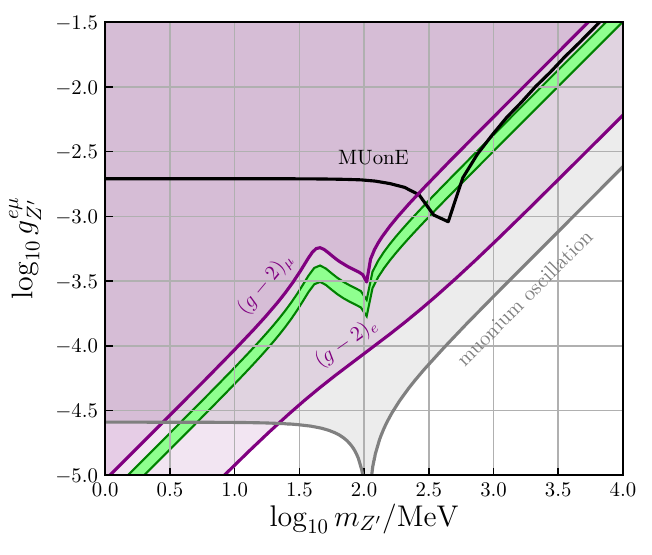}
  \caption{\label{fig:constraints_Z_s}Constraints on the tree-level $s$-channel $Z'$. The gray-shaded region is the muonium-antimuonium oscillation limit. The rest of the color coding is the same as in Fig.~\ref{fig:constraints_Z_t}.}
\end{figure}

The differential cross section of $\mu$--$e$ scattering in the presence of an $s$-channel $Z'$ boson reads:
\begin{eqnarray}
\frac{d\sigma}{dT} & \ = \ &
\left.\frac{d\sigma}{dT}\right|_{{\rm SM}}+\frac{1}{32\pi E_{\mu}^{2}m_{e}}\nonumber \\
 &  & \times\left[\frac{4e^{2} (g_{Z'}^{e\mu})^2 (s-m_{Z'}^{2})\left[2E_{\mu}m_{e}(E_{\mu}+2m_{\mu})+2m_{e}T^{2}-4E_{\mu}m_{e}T-(m_{e}+m_{\mu})^{2}T\right]}{T\left[(s-m_{Z'}^{2})^{2}+m_{Z'}^{2}\Gamma_{Z'}^{2}\right]}\right.\nonumber \\
 &  & \ \ \ \ \left.+\frac{8 (g_{Z'}^{e\mu})^4 m_{e}^{2}\left(E_{\mu}^{2}+2E_{\mu}(m_{\mu}-T)+T\left(m_{e}+m_{\mu}^{2}/m_{e}\right)+3m_{\mu}^{2}+2T^{2}\right)}{(s-m_{Z'}^{2})^{2}+m_{Z'}^{2}\Gamma_{Z'}^{2}}
\right],\label{eq:z-4}
\end{eqnarray}
where $\Gamma_{Z'} \simeq (g_{Z'}^{e\mu})^{2}m_{Z'}/8\pi$ is the width for the decay $Z'\rightarrow e\mu$.
With the cross section given in Eq.~(\ref{eq:z-4}) and using the
$\chi^{2}$-function from Eq.~\eqref{eq:chi2}, it is straightforward
to compute the sensitivity of MUonE for the $s$-channel $Z'$ mediator,
which is presented in Fig.~\ref{fig:constraints_Z_s} by the black
curve. As expected, when $m_{Z'}$ is close to the center-of-mass energy $\sqrt{s} \approx 406$ MeV, an $s$-channel resonance appears, leading to the dip at $\log_{10} (m_{Z'}/{\rm MeV})\simeq 2.6$. With the coupling $g_{Z'}^{e\mu}$, the $Z'$ boson contributes to the electron and muon $g-2$. When evaluating Eq.~(\ref{eq:g-2_v}), we take $\epsilon_{\ell'} = m_\mu/m_e \, (m_e/m_\mu)$ for the electron (muon) $g-2$. As in Fig.~\ref{fig:constraints_Z_t}, the shaded purple regions are excluded by the limits from the electron and muon $g-2$, and the green band indicates the parameter space for the muon $g-2$ discrepancy at the $1\sigma$ C.L.
There is also a resonance-like structure for the muon $g-2$ discrepancy and limit in Fig.~\ref{fig:constraints_Z_s}. This can be easily understood from Eq.~(\ref{eq:g-2_v}): the second term in the denominator is suppressed by the electron mass and can be neglected. In the limit of $m_{Z'} \to m_\mu$, we will have an extra factor of $(1-x)$ in the first term of the denominator, which will enhance the $Z'$ contribution to muon $g-2$ when we perform the integration over $x$.
When the $Z'$ boson has only the flavor-violating coupling $g_{Z'}^{e\mu}$, it can not induce either the process $e^+ e^- \to \gamma Z'$ or $e^+ e^- \to \mu^+ \mu^- Z' \to 4\mu$ in the BaBar experiment; therefore we do not have any limits from BaBar in this case.

For the $s$-channel $Z'$ boson, there is also the limit from muonium-antimuonium oscillation, where the $Z'$-induced mass splitting is given by~\cite{Bernstein:2013hba}
\begin{eqnarray}
\label{eqn:DeltaMZp}
|\Delta M|  \ = \
\frac{2 \alpha^3 |g_{Z'}^{e\mu}|^2 \mu^3}{\pi | (m_e + m_\mu)^2 - m_{Z'}^2 + i m_{Z'} \Gamma_{Z'} |} \,,
\end{eqnarray}
with $\mu = m_e m_\mu / (m_e + m_\mu)$ the effective mass. In Eq.~(\ref{eqn:DeltaMZp}) we have included the dependence on the $Z'$ width $\Gamma_{Z'}$. Then the muonium-antimuonium conversion probability due to the light $Z'$ boson is~\cite{Feinberg:1961zza}:
\begin{eqnarray}
{\cal P} \ = \ \frac{2 |\Delta M|^2}{\Gamma_\mu^2 + 4 |\Delta M|^2}\,,
\end{eqnarray}
where $\Gamma_\mu$ is the muon width.
The current limit of ${\cal P} <8.2 \times 10^{-11}$ from the MACS experiment~\cite{Willmann:1998gd} has excluded the gray-shaded region in Fig.~\ref{fig:constraints_Z_s}. When $m_{Z'}$ is close to $m_e + m_\mu$, the muonium-antimuonium oscillation probability is resonantly enhanced, and the constraint becomes significantly stronger, as shown by the dip in Fig.~\ref{fig:constraints_Z_s}.

From Fig.~\ref{fig:constraints_Z_s}, it is clear that the MUonE sensitivity, as well as the $1\sigma$-preferred region to explain the muon $(g-2)$ anomaly, is precluded by the electron $(g-2)$ constraint in this flavor-violating $Z'$ model. Thus, the MUonE measurement of $a_\mu^{\rm had}$ is immune to this new physics scenario.


\section{$L_{\mu}-L_{\tau}$ model}
\label{sec:loop}

In this section we consider one particular realistic model for the light $Z'$ boson, i.e.\ the $L_\mu - L_\tau$ model~\cite{He:1990pn,Foot:1990mn,He:1991qd}, which is one of the simplest models to account for the muon $g-2$ anomaly. Furthermore, it is attractive because
at zeroth order it predicts maximal atmospheric neutrino mixing, vanishing reactor mixing angle and a moderate neutrino mass hierarchy \cite{Ma:2001md,Heeck:2011wj}.
In this model a flavor-dependent leptonic $U(1)$ gauge symmetry is introduced to the $\mu$ and $\tau$ sector, and the beyond SM gauge couplings in the charged lepton sector are
\begin{equation}
{\cal L} \ \supset \ g_{Z'} Z'_\mu (\overline{\mu}\gamma^{\mu}\mu-\overline{\tau}\gamma^{\mu}\tau)\,.
\label{eq:z-3-1}
\end{equation}
Due to the absence of tree-level $Z'$ couplings to electrons, it can evade many stringent high and low energy constraints such as from LEP or the electron $g-2$.

\begin{figure}
  \centering
  \includegraphics[width=0.25\textwidth]{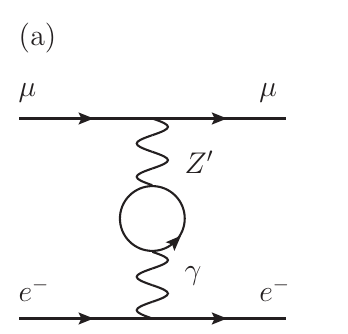}\hspace{1cm}
  \includegraphics[width=0.25\textwidth]{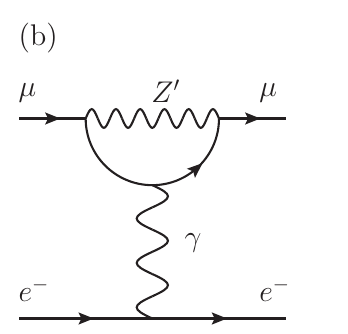}
  \caption{\label{fig:feyn-loop}
  Feynman diagrams of loop-level NP contributions to $\mu$--$e$ scattering in the $L_\mu - L_\tau$ model.
  }
\end{figure}

We point out that there is a loop-induced effective coupling to electrons (see Fig.~\ref{fig:feyn-loop}), so that it is similar to the case V(ii) with flavor-dependent flavor-conserving couplings, treated in Section~\ref{sub:Flavor-conserving}. There are however two differences: First, the effective coupling of $Z'$ to electrons is $q^2$-dependent. Second, there is a
second $Z'$-mediated diagram, shown in the right panel of Fig.~\ref{fig:feyn-loop}, where the $Z'$ boson is not even a $t$-channel mediator. Therefore, the $Z'$ boson in the $L_\mu - L_\tau$ model is  different from the case V(ii) in Section~\ref{sub:Flavor-conserving}.
Note further that diagrams with neutrinos in the loop, which would necessarily be present in case a gauge-invariant formulation is used, are highly suppressed by neutrino masses.



\begin{figure}
  \centering
  \includegraphics[height=0.45\textwidth]{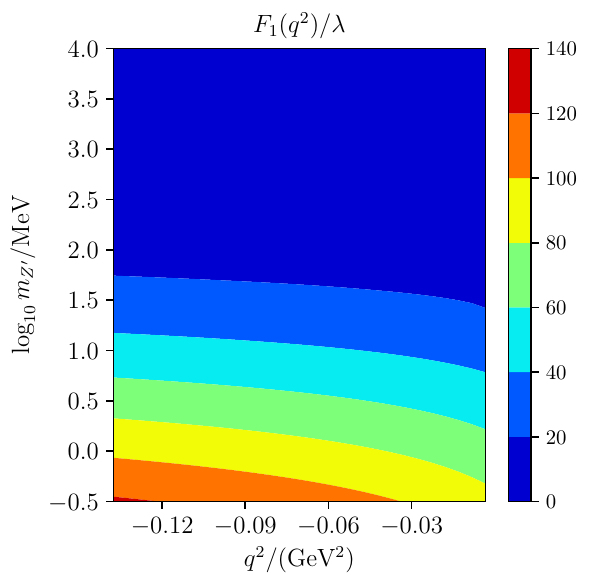} \ \ \ \includegraphics[height=0.45\textwidth]{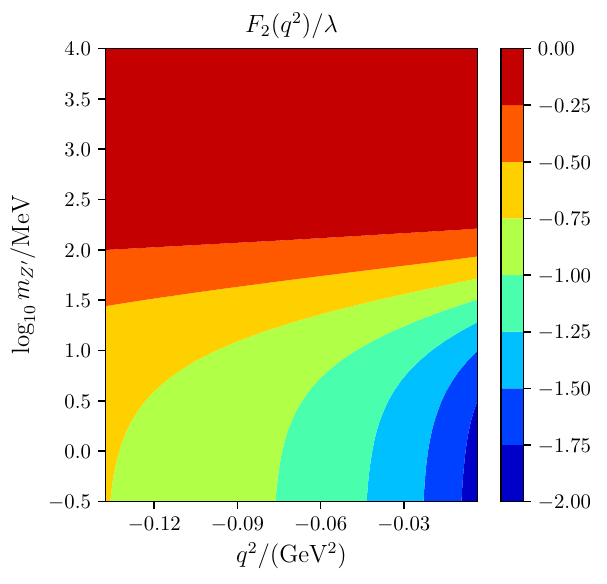}
  \caption{\label{fig:F1F2} Dependence of the form factors $F_{1}(q^{2})/\lambda$ (left) and $F_{2}(q^{2})/\lambda$ (right) on the $Z'$ mass and $q^2$ in the $L_{\mu}-L_{\tau}$ model, with $\lambda \equiv eg_{Z'}^{2}/16\pi^{2}$.}
\end{figure}


It is straightforward to calculate the loop diagrams in Fig.~\ref{fig:feyn-loop}, 
and the resulting differential cross section of $\mu$--$e$ scattering in the $L_{\mu}-L_{\tau}$ model reads
\begin{eqnarray}
\frac{d\sigma}{dT} & \ = \ &
e^{2}\left( F_{1}(q^{2})+e\right)^{2}
\frac{2E_{\mu}m_{e}(E_{\mu}-T)+T\left(m_{e}T-m_{e}^{2}-m_{\mu}^{2}\right)}{16\pi E_{\mu}^{2}m_{e}^{2}T^{2}}\nonumber \\
 && +e^{2} F_{2}^2(q^{2}) \frac{2E_{\mu}m_{e}(E_{\mu}-T)+m_{\mu}^{2}(T-2m_{e})}{32\pi E_{\mu}^{2}m_{e}m_{\mu}^{2}T}\nonumber \\
 &  & +e^{2} \left( F_{1}(q^{2})+e \right) F_{2}(q^{2})\frac{m_{e}-T}{8\pi E_{\mu}^{2}m_{e}T}~,\label{eq:z-5}
\end{eqnarray}
where $F_{1}(q^{2})$ and $F_{2}(q^{2})$ are momentum-dependent form factors for respectively the vector-current and dipole moment. The numerical dependence of  $F_{1}(q^2)$ and $F_2 (q^2)$ on the $Z'$ mass and $q^2$ are evaluated using {\tt Package-X}~\cite{Patel:2015tea, Patel:2016fam} and are shown respectively in the left and right panels of Fig.~\ref{fig:F1F2}. Here we have rescaled the form factors by the factor of $\lambda \equiv {g_{Z'}^{2}e}/{16\pi^{2}}$ such that the contours in Fig.~\ref{fig:F1F2} do not depend on the gauge coupling.

\begin{figure}
  \centering
  \includegraphics[width=8.25cm]{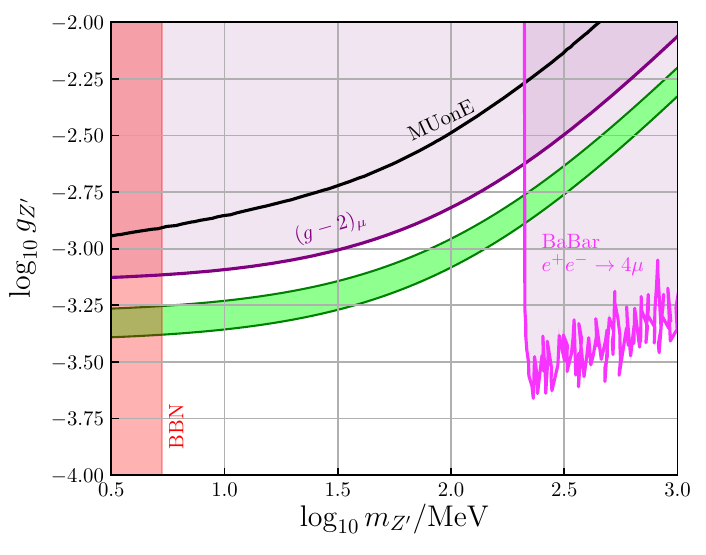} \caption{\label{fig:constraints-mutau}
  MUonE sensitivity and current constraints on $m_{Z'}$ and $g_{Z'}$ in the $L_{\mu}-L_{\tau}$ model.  The red shaded region is the BBN limit on $Z'$ mass and the color coding is the same as in Fig.~\ref{fig:constraints_Z_t}.  }
\end{figure}

Using the cross section in Eq.~(\ref{eq:z-5}) and the numerical
values of the form factors $F_{1,2}(q^{2})$, the MUonE sensitivity of the $L_\mu - L_\tau$ model is presented in Fig.~\ref{fig:constraints-mutau} by the black curve. As the $\mu$--$e$ scattering in the $L_\mu - L_\tau$ model is loop-suppressed, the MUonE sensitivity is weaker than that in case V(i) (see the left upper panel in Fig.~\ref{fig:constraints_Z_t}). The $Z'$ contribution to the muon $g-2$ is the same as that in case V(i) above (see the left upper panel in Fig.~\ref{fig:constraints_Z_t}), shown in Fig.~\ref{fig:constraints-mutau} as the purple curve, and precludes the MUonE sensitivity in the $L_\mu - L_\tau$ model. The $Z'$ contribution to the electron $g-2$ is loop-suppressed, hence much weaker, and not shown in Fig.~\ref{fig:constraints-mutau}.  The $Z'$ coupling to muons also induces the process $e^+ e^- \to \mu^+ \mu^- Z' \to 4\mu$ and thus gets constrained by the BaBar data~\cite{TheBABAR:2016rlg}.
This is shown as the pink shaded regions in Fig.~\ref{fig:constraints-mutau} which excludes any MUonE sensitivity for $m_{Z'} \gtrsim 200$ MeV.
Note that in a gauge-invariant $L_\mu - L_\tau$ realization, the invisible decay $Z'\rightarrow\nu\overline{\nu}$ also exists, which was not taken into account in Ref.~\cite{TheBABAR:2016rlg} searching for muonic dark forces. Including the partial invisible decay width, the bound is expected to be slightly weaker but it does not change our conclusions drawn from Fig.~\ref{fig:constraints-mutau}.
To be complete, we also consider the BBN limit on the $Z'$ boson in the $L_\mu - L_\tau$ model, which stems from the $Z'$ coupling to neutrinos. This requires that $m_{Z'} >5.3$ MeV~\cite{Kamada:2015era}, and is  indicated by the red shaded region in Fig.~\ref{fig:constraints-mutau}.

\section{Scalar mediators}
\label{sec:scalar}

\begin{figure}[!t]
  \centering

  \includegraphics[height=0.25\textwidth]{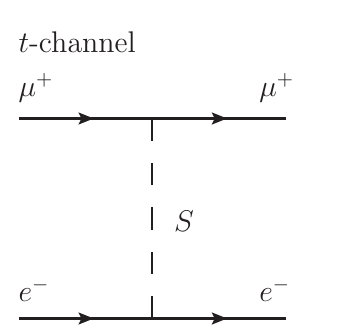}\hspace{1cm}
  \includegraphics[height=0.25\textwidth]{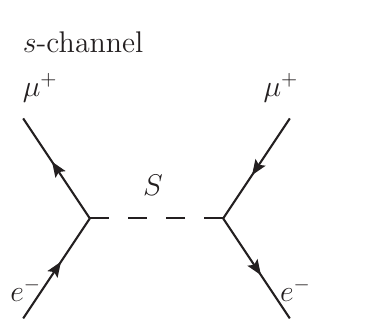}

  \caption{Feynman diagrams of $\mu$--$e$ scattering
  mediated by a flavor-conserving $S$ in the $t$-channel (left) or a flavor-changing $S$ in the $s$-channel (right).
  }
  \label{fig:feyn_tree_S}
  \end{figure}

In this section, we extend the analysis of the light $Z'$ boson cases in Section~\ref{sec:vector} to the light neutral scalar $S$ case. We will not consider any specific realistic model for the scalar $S$, rather perform a model-independent analysis for the MUonE sensitivities. As discussed in Section~\ref{sec:general_discuss}, the scalar mediator for the purpose of $\mu$--$e$ scattering can be either S(i) a light neutral scalar in the $t$-channel with flavor-conserving couplings $y_{ee,\,\mu\mu}$, or S(ii) a light neutral scalar in the $s$-channel with flavor-violating couplings $y_{e\mu}$.
The corresponding Feynman diagrams are shown in Fig.~\ref{fig:feyn_tree_S}. For simplicity we assume the light neutral scalar in both cases S(i) and S(ii) to be hadrophobic, i.e.\ with couplings to quarks and the SM Higgs vanishing or highly suppressed (for concrete examples, see e.g.~Refs.~\cite{Dev:2017ftk, Dev:2018upe}), such that we can avoid some potentially severe constraints including the current Higgs precision data and flavor-changing neutral currents in the quark sector. Again, as in the $Z'$ case, this implies that our conclusions are very conservative.


\subsection{Flavor-conserving couplings in the $t$-channel}
\label{sec:scalar1}

\begin{figure}
  \centering
  \includegraphics[width=0.98\textwidth]{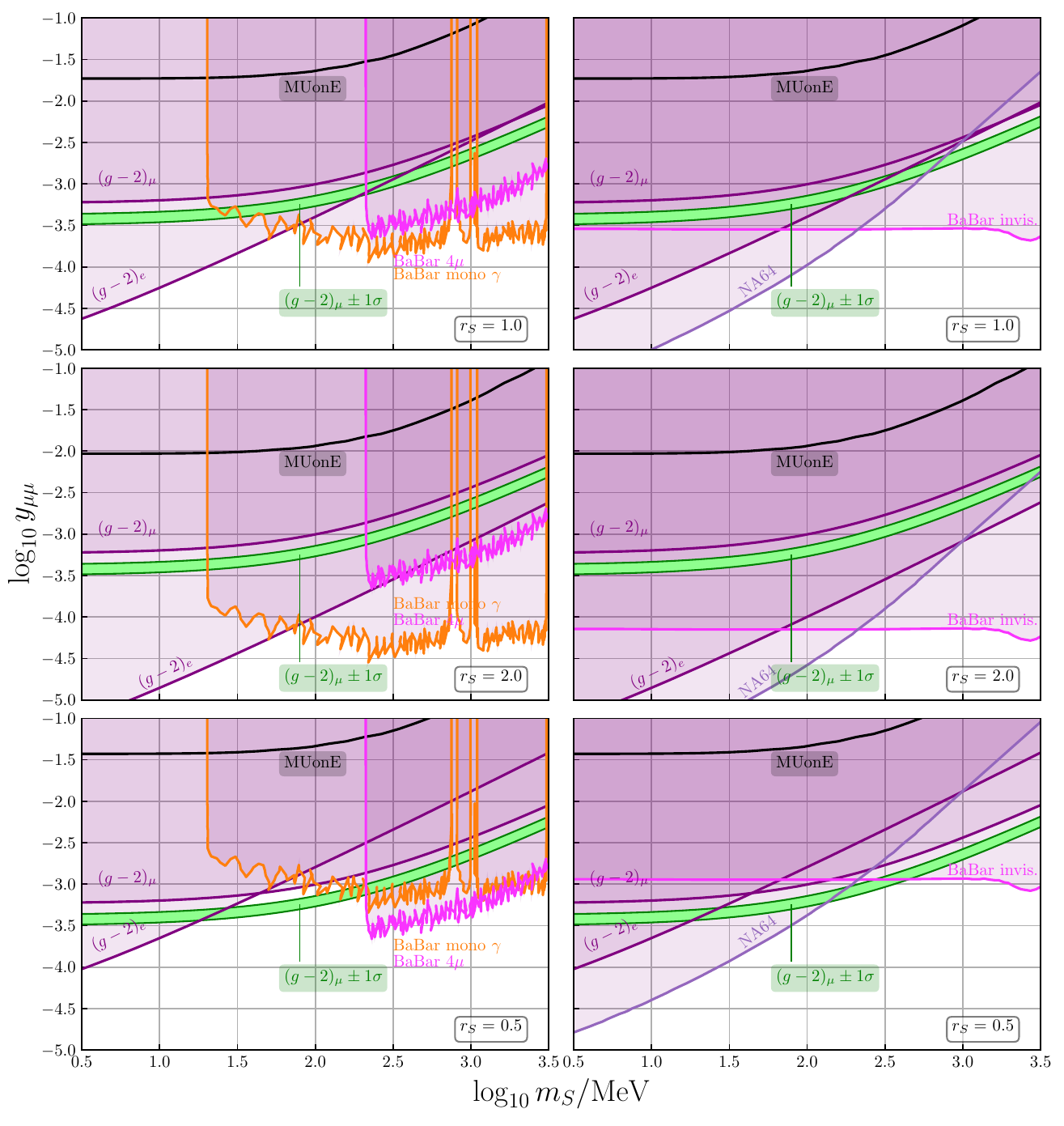}
  \caption{MUonE sensitivities on $m_S$ and $y_{\mu\mu}$ and current constraints on $S$ in the $t$-channel with the couplings in Eq.~(\ref{eq:s-FNU}) and for $r_S = 1$ (upper panel), 2 (middle panel) and 0.5 (lower panel), with $S$ decaying visibly (left panel) or invisibly (right panel), with the same color coding same as in Fig.~\ref{fig:constraints_Z_t}.  }
  \label{fig:scalar_r}
\end{figure}

Let us first consider a light scalar with flavor-conserving Yukawa couplings $y_{ee,\, \mu\mu}$ in Eq.~(\ref{eq:s-FNU}), which mediates the $t$-channel diagram in the left panel of Fig.~\ref{fig:feyn_tree_S}. The differential cross section in this case reads
\begin{eqnarray}
\label{eqn:cs:S1}
\frac{d\sigma}{dT} & \ = \ &
\frac{1}{4\pi (E_\mu^2-m_\mu^2)} \left[ \frac{e^4 \left( 2E_{\mu}m_{e}\left(E_{\mu}-T\right)
-T\left(m_{e}^{2}+m_{\mu}^{2}-m_{e}T\right) \right)}{4m_e^2 T^2} \right. \nonumber \\
&& \left. - \frac{e^2 |y_{ee} y_{\mu\mu}| m_\mu (2E_\mu -T)}{T(m_S^2 + 2 m_e T)}
+ \frac{|y_{ee} y_{\mu\mu}|^2 (2m_e+T) (2m_\mu^2 + m_e T)}{2(m_S^2 + 2 m_e T)^2}  \right],
\end{eqnarray}
where the first and third terms in the bracket are respectively the contributions from the photon [cf.~Section~\ref{sec:basic}] and the scalar $S$, and the second term is the interference term. The couplings of $S$ to electron ($y_{ee}$) and muon ($y_{\mu\mu}$) can be different, thus following the case V(ii) for the $Z'$ boson above we take three benchmark values for the Yukawa coupling ratio 
\begin{eqnarray}
r_S \ \equiv \ \sqrt{\frac{y_{ee}}{y_{\mu\mu}}} \ = \
\{ 1,\, 0.5,\, 2 \} \,.
\end{eqnarray}
Following the procedure in Section~\ref{sec:basic}, we obtain the prospects of the MUonE experiment for all three benchmark points from Eq.~(\ref{eqn:cs:S1}). The MUonE sensitivities of $m_{S}$ and $y_{\mu\mu}$ for $r_S = 1, \, 2$ and 0.5 are shown respectively in the upper, middle and lower panels of Fig.~\ref{fig:scalar_r} as the thick black curves. Comparing the three cases with different $r_S$ values, it is clear that a large value of $r_S$ will to some extent enhance the MUonE sensitivities, but this is not enough to overcome the current constraints. Note that the interference term between the SM and the scalar, i.e.\ the second term in Eq.~(\ref{eqn:cs:S1}), is helicity-suppressed by $m_{\mu}/E_{\mu}$ compared to the $Z'$ case [see Eq.~(\ref{eq:z-2})]. This is the reason why the MUonE sensitivities for the scalar case in Fig.~\ref{fig:scalar_r} are weaker than those for the $Z'$ case in Fig.~\ref{fig:constraints_Z_t}.

If the scalar $S$ couples only to electron and muon as given by the Lagrangian in Eq.~(\ref{eq:s-FNU}), the limits on the Yukawa couplings $|y_{ee}|$ and $|y_{\mu\mu}|$ are mainly from precision measurements of leptonic processes. As here $S$ does not have the LFV coupling like $y_{e\mu}$, the LFV limits~\cite{Dev:2017ftk} such as those from $\mu \to e\gamma$, $\mu \to eee$ and muonium-antimuonium oscillation do not apply. The single coupling $y_{ee}$ could induce M{\o}ller scattering $e^- e^- \to e^- e^-$, with $S$ playing the role of a $t$-channel mediator, and thus gets constrained by the future high-precision MOLLER experiment~\cite{Benesch:2014bas} (see also~\cite{Dev:2018sel}). However, without involving $y_{\mu\mu}$, the MOLLER constraint on $|y_{ee}|$ can not be used to set limits on the combination $\sqrt{|y_{ee} y_{\mu\mu}|}$.
We find that the most stringent constraints come from electron and muon $g-2$ and the BaBar experiment.

Constraints from electron and muon $g-2$ can be computed according to Eq.~(\ref{eq:g-2_s}) with $\ell'=\ell$ and $\epsilon_{\ell'}=1$. It is important to note that the MUonE experiment is only sensitive to the product $y_{ee}y_{\mu\mu}$, while electron and muon $g-2$ only provide constraints on $y_{ee}$ and $y_{\mu\mu}$, respectively. By combining electron and muon $g-2$ constraints, one can get a constraint on $y_{ee}y_{\mu\mu}$. As shown in Fig.~\ref{fig:scalar_r}, for all three cases of $r_S = 1,\, 2$ and 0.5, the electron and muon $g-2$ limits have fully precluded the parameter space of the MUonE sensitivities. Actually this is also true for any value of $r_S$, and we can now draw the conclusion that NP in this scenario cannot affect the MUonE measurement at any significance level.

For the sake of completeness, we add also the limits from BaBar and the beam-dump experiment NA64, although this does not change our conclusion. As for the $Z'$ case, the BaBar limits depend on how the scalar $S$ decays. Given the couplings in Eq.~(\ref{eq:s-FNU}), the scalar $S$ can decay into electrons and muons, i.e.\ $S \to e^+ e^- ,\, \mu^+ \mu^-$. It  might also decay into invisible particles like neutrinos or some light dark sector particles. The BaBar and NA64 limits on the visibly and invisibly decaying $S$ are completely analogous to the $Z'$ boson case in Section~\ref{sub:Flavor-conserving}, and  these limits are shown respectively in the left and right panels of Fig.~\ref{fig:scalar_r}, with the same color coding as in Fig.~\ref{fig:constraints_Z_t}.

\subsection{Flavor-changing couplings in the $s$-channel}
\label{sec:scalar2}

\begin{figure}
  \centering
  \includegraphics[width=7.5cm]{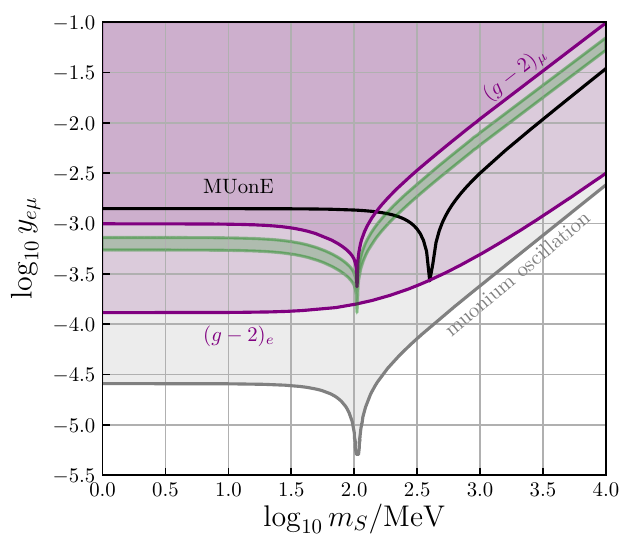}
  \caption{
  MUonE sensitivity and current constraints on $m_S$ and $y_{e\mu}$ for $S$ in the $s$-channel.  The gray shaded region is the muonium-antimuonium oscillation limit. The rest of the color coding is the same as in Fig.~\ref{fig:constraints_Z_t}.}
  \label{fig:scalar2}
\end{figure}

As previously discussed in Section~\ref{sec:general_discuss}, for a light flavor-changing neutral scalar, only the $y_{e\mu}$ coupling is relevant to the MUonE experiment. It contributes to the $\mu$--$e$ scattering via the $s$-channel, as shown in the right panel of Fig.~\ref{fig:feyn_tree_S}. In the absence of other Yukawa interactions such as $y_{ee}$ and $y_{\mu\mu}$, the LFV decays $\mu \to e\gamma$ and $\mu \to eee$ do not receive any new contribution from $S$. The differential cross section reads
\begin{eqnarray}
\label{eq:dsigmadT:scalar2}
\frac{d\sigma}{dT} & \ = \ &
\frac{1}{16\pi m_e^2 E_\mu^2} \left[
\frac{e^4 \left( 2E_{\mu}m_{e}\left(E_{\mu}-T\right)
-T\left(m_{e}^{2}+m_{\mu}^{2}-m_{e}T\right) \right)}{T^2} \right. \nonumber \\
&& + \frac{e^2 |y_{e\mu}|^2 m_e \left( 2 m_e E_\mu (m_e-E_\mu) + T (m_e+m_\mu)^2 \right) (s - m_s^2)}{T \left[ (s - m_s^2)^2 + m_s^2 \Gamma_s^2 \right]} \nonumber \\
&& \left. +
\frac{2 |y_{e\mu}|^4 m_e^3 (E_\mu-m_\mu)^2}{(s - m_s^2)^2 + m_s^2 \Gamma_s^2}  \right] ,
\end{eqnarray}
where the second and third terms in Eq.~(\ref{eq:dsigmadT:scalar2}) are the NP contributions. If the scalar $S$ mass lies in the sub-GeV scale, it might be produced resonantly in the MUonE experiment, and we have included the width $\Gamma_S \simeq |y_{e\mu}|^2 m_S/8\pi$ in Eq.~(\ref{eq:dsigmadT:scalar2}). The MUonE sensitivity on this scenario is presented in Fig.~\ref{fig:scalar2} as the black line. Again, as a consequence of the resonance effect, there is a dip at the center-of-mass energy $\sqrt{s} \simeq 406$ MeV.

Based on Eq.~(\ref{eq:g-2_s}) (note that in this case $\ell \neq \ell'$, $\epsilon_e = m_e/m_\mu$ and $\epsilon_\mu = m_\mu/m_e$), the electron and muon $g-2$ give rise to stringent constraints on the scalar mass $m_S$ and the coupling $y_{e\mu}$.
It is clear from Fig.~\ref{fig:scalar2} that the MUonE prospect has been precluded by the electron $g-2$ limit. As in the $Z'$ case (see Section~\ref{sub:Flavor-changing}), there is also a resonance structure for the muon $g-2$  in the limit of $m_S \to m_\mu$, as there will be an extra factor of $(1-x)$ in the denominator of Eq.~(\ref{eq:g-2_s}).

In addition to the electron and muon $g-2$ limits, the $y_{e\mu}$ coupling also contributes to the muonium-antimuonium oscillation, and the resulting  constraint turns out to be more stringent. The $S$-induced mass splitting is quite similar to the $Z'$ case in Eq.~(\ref{eqn:DeltaMZp})~\cite{Clark:2003tv}:
\begin{eqnarray}
\label{eqn:DeltaM}
|\Delta M|  \ = \
\frac{2 \alpha^3 |y_{e\mu}|^2 \mu^3}{\pi | (m_e + m_\mu)^2 - m_{S}^2 + i m_S \Gamma_S |} \,,
\end{eqnarray}
with $\Gamma_S$ the width of $S$. The resultant limit on $m_S$ and $y_{e\mu}$ is indicated by the gray-shaded region in Fig.~\ref{fig:scalar2}. As for the $Z'$ case, when $m_S$ is close to $m_e + m_\mu$, the muonium-antimuonium constraint is resonantly enhanced. The $y_{e\mu}$ coupling of $S$ will also induce the scattering $e^+ e^- \to \mu^+ \mu^-$ at LEP, however, as for the case of $t$-channel light scalar in Section~\ref{sec:scalar1}, the limit from LEP data is much weaker, $y_{e\mu} \lesssim 0.041$, and thus not shown in Fig.~\ref{fig:scalar2}.

\section{Conclusion \label{sec:Conclusion}}

The MUonE experiment is of great importance in reducing the current theoretical uncertainty in the hadronic contribution to the muon anomalous magnetic moment. Since there may be new physics related to the muon $g-2$ anomaly, it is necessary to study whether the MUonE measurement of $a^{\rm had}_{\mu}$ could also be affected by such NP. In this paper, we have considered generic light scalar and vector bosons that  couple to electron and muon, and comprehensively studied the corresponding effects on the MUonE experiment. Our conclusions, drawn from very conservative assumptions, can be summarized as follows:
\begin{itemize}
  \item For NP scenarios involving the light scalar $S$, MUonE is very safe from their potential influence because the NP parameter space accessible to MUonE has been fully precluded by  the stringent limits from electron and muon $g-2$, BaBar and NA64, which are typically one order of magnitude stronger than the MUonE sensitivities;  see Figs.~\ref{fig:scalar_r} and \ref{fig:scalar2}.
  \item If a $Z'$ boson has flavor-conserving flavor-universal couplings to electron and muon, i.e.\ $g_{Z'}^e = g_{Z'}^{\mu}$, or has only flavor-changing coupling $g_{Z'}^{e\mu}$, then the corresponding parameter space accessible to MUonE has also been excluded by the  limits from electron and muon $g-2$, BaBar and NA64; see the two upper panels in
  Fig.~\ref{fig:constraints_Z_t} and Fig.\ \ref{fig:constraints_Z_s}. For the case with flavor-non-universal couplings $g_{Z'}^e \neq g_{Z'}^{\mu}$, almost the full parameter space has been excluded, except for a narrow window close to BaBar limits, as shown in the lower panels of Fig.~\ref{fig:constraints_Z_t}. We have chosen typical benchmark points in this regime, and demonstrated that they would have only a small effect less than around $1\sigma$ on MUonE's determination of $a_\mu^{\rm had}$, as shown in  Fig.~\ref{fig:a_had}.
  \item In addition to the model-independent analysis above, we have also studied a realistic $L_\mu - L_\tau$ model, as it is very different from the scenarios above and provides a very good example for loop-level NP contribution to $\mu$--$e$ scattering (see Fig.~\ref{fig:feyn-loop}), since the light $Z'$ boson in this case does not couple directly to electron at tree-level. However, the corresponding MUonE sensitivity has been excluded by muon $g-2$ limit and BaBar data (see Fig.~\ref{fig:constraints-mutau}).
\end{itemize}

We thus conclude that the MUonE measurement of $a^{\rm had}_{\mu}$ is invulnerable to new physics that might be responsible for the muon $g-2$ anomaly. If the future MUonE measurement of $a^{\rm had}_{\mu}$ is not consistent with the current theoretical estimation given by Eq.~\eqref{eq:a-had}, it is very unlikely due to NP contributions.







{{\bf Note added:} After completion of this work, we became aware of Ref.~\cite{Masiero:2020vxk} which studies the MUonE sensitivity
to both heavy and light mediators, and reaches the same conclusion as ours.}

\begin{acknowledgments}
We thank Julian Heeck for helpful discussions on $L_{\mu}-L_{\tau}$ and muon $g-2$. We are also very grateful to Antonio Masiero, Paride Paradisi and Massimo Passera for cross-checking our results with those in Ref.~\cite{Masiero:2020vxk} and for correcting our Eqs.~(\ref{eq:SM-a}) and (\ref{eqn:cs:S1}). X-J.X. would like to thank the organizers of the NTN NSI Workshop at Washington University in St. Louis, where this work was initiated. B.D. and X-J.X. would like to thank the Fermilab Theory Group for hospitality during a summer visit, where part of this work was done. Y.Z. would like to thank Graziano Venanzoni for discussions about the MUonE experiment at the ICHEP 2018 conference in Seoul, Korea, also to Yuber Perez-Gonzalez for the inspiring discussion on the direct production of light particles in the MUonE experiment, and is grateful to the Center for Future High Energy Physics, Institute of High Energy Physics, Chinese Academy of Sciences for generous hospitality where part of this work was done.  The work of B.D. and Y.Z. is supported by the US Department of Energy under Grant No. DE-SC0017987 and the Neutrino Theory Network Program under Grant No. DE-AC02-07CH11359. W.R. is supported by the DFG with grant RO 2516/7-1 in the Heisenberg program.

\end{acknowledgments}

\bibliographystyle{JHEP}
\bibliography{ref}

\end{document}